\documentclass[traditabstract]{aa}

\usepackage[pdftex]{graphicx}
\usepackage{amsmath}
\usepackage[mathscr]{eucal}
\usepackage{amssymb,amsfonts}
\usepackage{natbib}

\usepackage{xcolor}

\bibpunct{(}{)}{;}{a}{}{,} 

\def\deg{\ifmmode^\circ\else$^\circ$\fi}
\def\pdeg{\ifmmode $\setbox0=\hbox{$^{\circ}$}\rlap{\hskip.11\wd0 .}$^{\circ}
          \else \setbox0=\hbox{$^{\circ}$}\rlap{\hskip.11\wd0 .}$^{\circ}$\fi}
\def\arcs{\ifmmode {^{\scriptscriptstyle\prime\prime}}
          \else $^{\scriptscriptstyle\prime\prime}$\fi}
\def\arcm{\ifmmode {^{\scriptscriptstyle\prime}}
          \else $^{\scriptscriptstyle\prime}$\fi}
\newdimen\sa  \newdimen\sb
\def\parcs{\sa=.07em \sb=.03em
     \ifmmode \hbox{\rlap{.}}^{\scriptscriptstyle\prime\kern -\sb\prime}\hbox{\kern \sa}
     \else \rlap{.}$^{\scriptscriptstyle\prime\kern -\sb\prime}$\kern -\sa\fi}
\def\parcm{\sa=.08em \sb=.03em
     \ifmmode \hbox{\rlap{.}\kern\sa}^{\scriptscriptstyle\prime}\hbox{\kern-\sb}
     \else \rlap{.}\kern\sa$^{\scriptscriptstyle\prime}$\kern-\sb\fi}
\def\microns{\ifmmode \,\mu$m$\else \,$\mu$m\fi}

\definecolor{Blue}{rgb}{0.,0.,1.}
\definecolor{Red}{rgb}{1., 0., 0.}
\newcommand\T{\rule{0pt}{2.6ex}}
\newcommand\B{\rule[-1.2ex]{0pt}{0pt}}

\newcommand{\degres}{\ensuremath{^\circ}}

\newcommand{\InsLERMA}{LERMA, CNRS, Observatoire de Paris, 61 Avenue de l'Observatoire, 75014 Paris, France}

\newcommand{\InsCESR}{CESR, CNRS, 9 av. du colonel Roche, BP44346, 31038 Toulouse Cedex 4, France}

\newcommand{\InsCARDIFF}{Astronomy and Instrumentation Group, Cardiff University, Cardiff, Wales}

\newcommand{\InsManchester}{The University of Manchester, JCBA, School of Physics and Astronomy, Manchester M13 9PL, UK}

\newcommand{\InsCaltech}{Observational Cosmology, California Institute of Technology, Mail code:367-17, Pasadena, CA 91125, USA}

\newcommand{\InsIAS}{IAS, Institut d'Astrophysique Spatiale, CNRS Universit\'e Paris 11, B\^atiment 121, 91405 Orsay, France}

\newcommand{\InsLAL}{LAL, Laboratoire de l'Acc\'elerateur Lin\'eaire, CNRS Universit\'e Paris 11, B\^atiment 200, Orsay, France}

\newcommand{\InsLPSC}{LPSC, Laboratoire de Physique Subatomique et Cosmologie, CNRS, Grenoble, France}

\newcommand{\InsLAOG}{LAOG, Laboratoire d'Astrophysique Observatoire de Grenoble, CNRS, Grenoble, France}

\newcommand{\InsRoma}{Dipartimento di Fisica, Universit\'a di Roma ``La Sapienza'', 00185 Roma, Italy}

\newcommand{\InsCEA}{CEA, Service de Physique des Particules, Saclay, France}

\newcommand{\InsAPC}{APC, Astroparticule et Cosmologie, Universit\'e
  Paris Diderot, B\^atiment Condorcet, 10 rue Alice Domon et L\'eonie
  Duquet, F-75205 Paris Cedex 13, France}

\newcommand{\InsPrinceton}{Department of Physics, Princeton University, Princeton, NJ 08544}

\newcommand{\InsOSL}{Optical Science Laboratory, University College London, Gower Street, WC1E 6BT London, UK}

\newcommand{\InsJPL}{Jet Propulsion Laboratory, California Institute of Technology, Pasadena, CA 91109, USA}

\newcommand{\InsESAC}{European Space Astronomy Centre, P.O.Box 78, 28691 Villanueava de la Ca\~{n}ada (Madrid), Spain}

\newcommand{\AetAInstituteCount}[2]{
  \setcounter{#1}{#2}
  \addtocounter{#1}{-1}
  \refstepcounter{#1}
  \label{#1}
}

\begin{document}

\title{Planck pre-launch status: High Frequency Instrument polarization calibration}
\author{C. Rosset\inst{\ref{InsLALcounter}, \ref{InsAPCcounter}} \and 
M. Tristram\inst{\ref{InsLALcounter}} \and 
N. Ponthieu\inst{\ref{InsIAScounter}} \and 
P. Ade\inst{\ref{InsCARDIFFcounter}} \and
A. Catalano\inst{\ref{InsLERMAcounter},\ref{InsAPCcounter}} \and
L. Conversi\inst{\ref{InsESACcounter}} \and
F. Couchot\inst{\ref{InsLALcounter}} \and
B. P. Crill\inst{\ref{InsCaltechcounter},\ref{InsJPLcounter}} \and
F.-X. D\'esert\inst{\ref{InsLAOGcounter}} \and
K. Ganga\inst{\ref{InsAPCcounter}} \and
M. Giard\inst{\ref{InsCESRcounter}} \and
Y. Giraud-H\'eraud\inst{\ref{InsAPCcounter}} \and
J. Ha\"{i}ssinski\inst{\ref{InsLALcounter}} \and
S. Henrot-Versill\'e\inst{\ref{InsLALcounter}} \and
W. Holmes\inst{\ref{InsJPLcounter}} \and
W. C. Jones\inst{\ref{InsCaltechcounter},\ref{InsJPLcounter},\ref{InsPrincetoncounter}} \and
J.-M. Lamarre\inst{\ref{InsLERMAcounter}} \and
A. Lange\inst{\ref{InsCaltechcounter},\ref{InsJPLcounter}}$^\dag$ \and
C. Leroy\inst{\ref{InsIAScounter},\ref{InsCESRcounter}} \and
J. Mac\'{i}as-P\'erez\inst{\ref{InsLAOGcounter}} \and
B. Maffei\inst{\ref{InsManchestercounter}} \and
P. de Marcillac\inst{\ref{InsIAScounter}} \and
M.-A. Miville-Desch\^enes\inst{\ref{InsIAScounter}} \and
L. Montier\inst{\ref{InsCESRcounter}} \and
F. Noviello\inst{\ref{InsIAScounter}} \and
F. Pajot\inst{\ref{InsIAScounter}} \and
O. Perdereau\inst{\ref{InsLALcounter}} \and
F. Piacentini\inst{\ref{InsRomacounter}} \and
M. Piat\inst{\ref{InsAPCcounter}} \and
S. Plaszczynski\inst{\ref{InsLALcounter}} \and
E. Pointecouteau\inst{\ref{InsCESRcounter}} \and
J.-L. Puget\inst{\ref{InsIAScounter}} \and
I. Ristorcelli\inst{\ref{InsCESRcounter}} \and
G. Savini\inst{\ref{InsCARDIFFcounter},\ref{InsOSLcounter}} \and
R. Sudiwala\inst{\ref{InsCARDIFFcounter}} \and
M. Veneziani\inst{\ref{InsAPCcounter},\ref{InsRomacounter}} \and
D. Yvon\inst{\ref{InsCEAcounter}}
}

\institute{
    \InsLAL 
    \AetAInstituteCount{InsLALcounter}{\value{inst}}
    \and \InsIAS 
    \AetAInstituteCount{InsIAScounter}{\value{inst}}
   \and \InsCARDIFF 
    \AetAInstituteCount{InsCARDIFFcounter}{\value{inst}}
    \and \InsLERMA 
    \AetAInstituteCount{InsLERMAcounter}{\value{inst}}
    \and \InsAPC  
    \AetAInstituteCount{InsAPCcounter}{\value{inst}}
    \and \InsCaltech 
    \AetAInstituteCount{InsCaltechcounter}{\value{inst}}
    \and \InsLAOG 
    \AetAInstituteCount{InsLAOGcounter}{\value{inst}}
    \and \InsCESR 
    \AetAInstituteCount{InsCESRcounter}{\value{inst}}
    \and \InsJPL 
    \AetAInstituteCount{InsJPLcounter}{\value{inst}}
    \and \InsManchester 
    \AetAInstituteCount{InsManchestercounter}{\value{inst}}
    \and \InsLPSC 
    \AetAInstituteCount{InsLPSCcounter}{\value{inst}}
   \and \InsRoma 
    \AetAInstituteCount{InsRomacounter}{\value{inst}}
    \and \InsCEA 
    \AetAInstituteCount{InsCEAcounter}{\value{inst}}
    \and \InsPrinceton 
    \AetAInstituteCount{InsPrincetoncounter}{\value{inst}}
    \and \InsOSL 
    \AetAInstituteCount{InsOSLcounter}{\value{inst}}
   \and \InsESAC
    \AetAInstituteCount{InsESACcounter}{\value{inst}}
} 

\date{Accepted April, 7$^{\mbox{\tiny th}}$ 2010}

\abstract {
The High Frequency Instrument of Planck will map the entire sky in the
millimeter and sub-millimeter domain from 100 to 857~GHz with
unprecedented sensitivity to polarization ($\Delta P/T_{\mbox{\tiny cmb}} \sim 4\cdot
10^{-6}$ for $P$ either $Q$ or $U$ and $T_{\mbox{\tiny cmb}}\simeq
2.7\,$K) at 100, 143, 217 and 353 GHz.  It
will lead to major improvements in our understanding of the Cosmic Microwave Background
anisotropies and polarized foreground signals. Planck will
make high resolution measurements of the $E$-mode spectrum (up to $\ell \sim
1500$) and will also play a prominent role in the search for 
the faint imprint of primordial gravitational waves on
the CMB polarization.\\
This paper addresses the effects of calibration of both temperature
(gain) and polarization (polarization efficiency and detector
orientation) on polarization measurements. The specific requirements
on the polarization parameters of the instrument are set and we
report on their pre-flight measurement on HFI bolometers.\\
We present a semi-analytical method that exactly accounts for the
scanning strategy of the instrument as well as the combination of
different detectors.  We use this method to propagate errors through
to the CMB angular power spectra in the particular case of Planck-HFI,
and to derive constraints on polarization parameters.\\ 
We show that in order to limit the systematic error to 10\% of the
cosmic variance of the $E$-mode power spectrum, uncertainties in gain,
polarization efficiency and detector orientation must be below 0.15\%,
0.3\% and 1\deg\ respectively. Pre-launch ground measurements reported
in this paper already fulfill these requirements.} 

\authorrunning{C. Rosset, M. Tristram, N. Ponthieu {\it et al}}
\titlerunning{Planck-HFI: Polarization Calibration}

\maketitle

\section{Introduction}

The Planck satellite, launched on May 14th, 2009, will map the whole
sky in the range 30--857~GHz. One of the most exciting challenges for
Planck is to measure the polarization anisotropies of the Cosmic
Microwave Background (CMB), which offers a unique way to constrain the
energy scale of inflation.

CMB polarization can be decomposed into modes of even-parity
($E$-mode) and odd-parity ($B$-mode). Gravitational waves generated
during inflation (hereafter ``primordial'' gravitational waves) create
$B$-modes with a specific angular power spectrum, whose amplitude is
related to the energy scale of inflation. A detection of these
``primordial'' $B$-modes would therefore provide the first measure of
the energy scale of inflation.

$E$-modes were first detected by \textsc{Dasi} in 2002, followed by
other ground and balloon-borne experiments \citep{Kovac:2002lt,
  Readhead:lk, Wu:ni, Montroy:2006vk,
  QUaD-collaboration:-C.Pryke:2008wo} covering a few percent of the
sky. These detections are complemented by the \textsc{Wmap} satellite
observations of the whole sky \citep{Page:2007rw}. All these
measurements have confirmed the existence of an $E$-mode polarization
compatible with the $\Lambda CDM$ model, and are compatible with a
$B$-mode polarization of zero. The tensor-to-scalar ratio $r$
parametrizes the amplitude of $B$-mode polarization. The most stringent
upper limit on $r$ is obtained by \cite{Komatsu:2009zs}, combining
\textsc{Wmap} measurements of TT, TE and EE power spectra with Baryon
Acoustic Oscillations and supernovae data. They obtain $r<0.22$ if the
scalar spectral index $n_S$ is constant, or $r<0.55$ if a running
spectral index is allowed.

Planck has been designed to map the $E$-mode of polarization with high
precision and good control over the polarization foreground
contamination up to multipoles as large as $\ell \sim 1500$. Planck
may also detect the $B$-mode polarization anisotropies, if tensor
modes contribute at a level of a few percent or more of the amplitude
of the scalar modes \citep{Efstathiou:2009kx}. However, various
instrumental systematic effects, induced by error on the knowledge of
detector characteristics, may alter these measurements.  Most of the
properties of the detectors, such as the gain, time constant, bandpass
and beam, are independent of the sensitivity to linear
polarization. These properties are described in detail in companion
papers \citep{Pajot:2010fr,Lamarre:2010fr,Tauber:2010ay,Maffei:2010ly}.
In this paper, we study the systematic effects induced by
uncertainties in temperature and polarization calibration (gains,
polarization efficiencies and orientations) on Stokes parameters and
$E$ and $B$-mode power spectra.  We also report on the ground
calibration of the polarization efficiencies and orientations of High
Frequency Instrument (HFI) detectors.
A study of polarization systematics for the Low Frequency Instrument
(LFI) of Planck is presented in \cite{leahy:2010}.

The paper is organized as follows. In Sect.~\ref{sec:detfp}, we
present the Polarization Sensitive Bolometers (PSBs) used by the
Planck HFI and the layout of the focal plane. Section~\ref{sec:photo}
gives the generic expression of the polarized photometric equation and
introduces the polarization-related systematic effects discussed in
Sect.~\ref{sec:syste}. In Sect.~\ref{sec:method}, we describe a
semi-analytical method to propagate uncertainties on temperature and
polarization calibration of detectors up to angular power spectra
while exactly accounting for the scanning strategy and the combination
of multiple detectors. We apply this method to the Planck HFI
in Sect.~\ref{sec:hfi} and derive requirements on the knowledge of
these parameters. Finally, Sect.~\ref{sec:ground_cal} describes the
procedure used to measure polarization parameters of the detectors on
ground and compares them to the requirements derived in the previous
section.

\section{Detectors and focal plane layout}
\label{sec:detfp}

\begin{figure}[tb]
\begin{center}
\includegraphics[width=6cm]{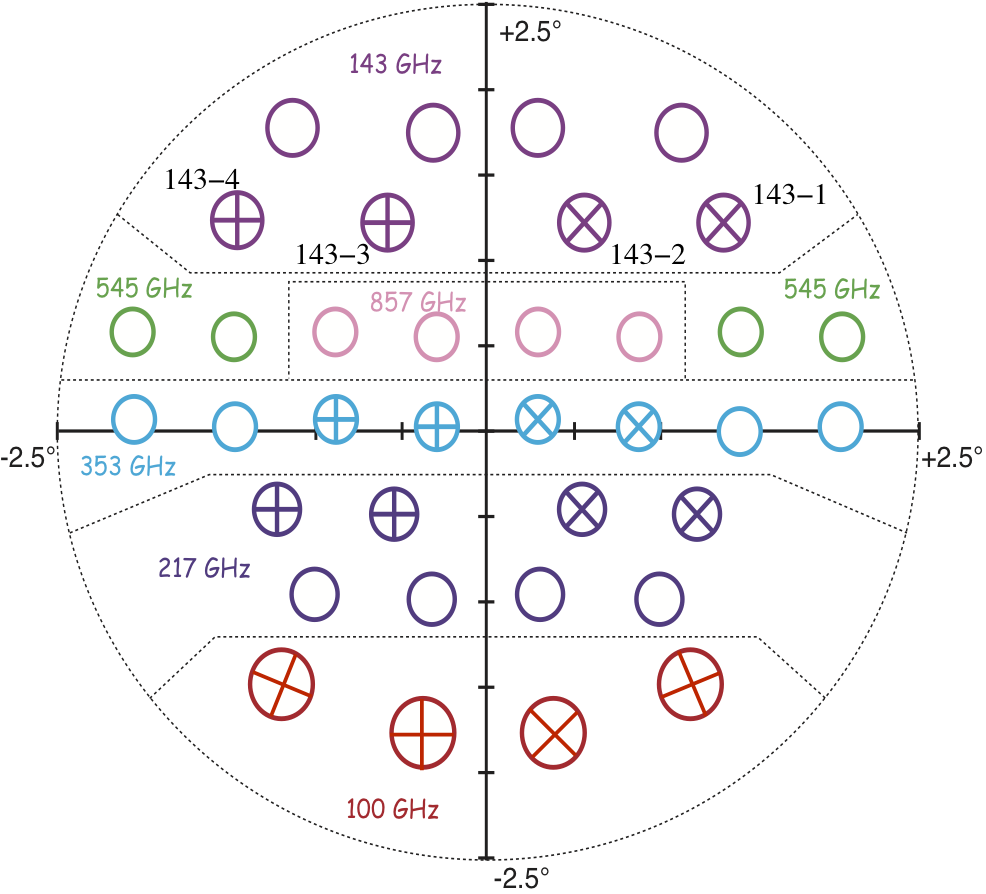}
\caption{\label{fig:planck-fplayout} Sky projection of the Planck HFI
  focal plane. The crosses symbolize the Polarization Sensitive
  Bolometers and indicate the orientation of the two linear
  polarization measured in each horn. The scanning direction is
  horizontal in this sketch, so that PSB pairs at same frequency
  follow the same track on the sky.}
\end{center}
\end{figure}

HFI uses bolometric detectors cooled to 100\,mK to measure
millimeter-wave radiation. They comprise a micro-mesh absorber in a
form resembling a spider web to reduce cosmic ray interactions \citep[hence the
  name Spider Web Bolometer or SWB, see][]{Bock:1995ty,Yun:2004sf},
heated by ohmic power dissipation, and a neutron transmutation doped
(NTD) germanium thermistor that measures the temperature
variation. Polarization is measured with specifically designed
polarization sensitive bolometers \citep[PSBs, see][]{Jones:2003mz},
composed of a pair of bolometers that couple to orthogonal linear
polarizations, allowing the measurement of $I$ and (local) $Q$ Stokes
parameters (respectively the sum and difference of the signals of the
two bolometers). The
SWBs are only slightly sensitive to polarization, and PSBs do not perfectly
reject the cross-polarization component. We define precisely the
cross-polarization leakage in the next section. The HFI focal plane is
composed of 20~SWBs and 16~PSB pairs, i.e., 32~polarization sensitive
bolometers (see Fig.~\ref{fig:planck-fplayout}). The PSB pairs are
grouped in pairs rotated by $45^\circ$ and following the same track
on the sky, with the angular separation between associated pairs ranging
from 0\pdeg5 to 2\pdeg5. Thus, the difference signal within one pair
measures Stokes $Q$ (in some local reference frame) while the difference
signal within the other pair measures Stokes $U$. Both pairs allow
measurements of the total intensity through the sum of signals. This
layout was chosen in order to minimize the noise on the Stokes
parameters and their correlation \citep{Couchot:1999jq}.

The satellite scans the sky by spinning at 1~rpm. The spin axis
remains within 7\pdeg5 of the anti-solar direction \citep[for a
  detailed presentation of the Planck scanning strategy, see
][]{Tauber:2010}. The angle between the spin axis and the line of
sight is $85\deg \pm 2\pdeg5$ depending on detector position in focal
plane. The ecliptic pole regions are thus much more covered than the
equatorial region, both in terms of number of hits per pixel and in
different observation orientations. This means that around the
ecliptic poles, each detector observes the sky with several focal
plane orientations and hence measures $I$, $Q$ and $U$. In contrast,
in the equatorial region, at least three detectors must be combined to
obtain the polarization signal. This is very different from currently
designed ground or balloon-borne experiments in which the Stokes
parameters can be measured using a single detector. This impacts
the propagation of errors, as discussed in detail in
Sect.~\ref{sec:method}.

\section{Polarized photometric equation}
\label{sec:photo}

In this section, we derive the expression for the power received by a
PSB. Following Jones's notation \citep{Jones:1941}, the polarization
state of a plane wave can be described by its transverse electric
field $\mathbf{e}=(E_x,E_y)$, where $E_x$ and $E_y$ are complex
amplitudes. The transmission through an instrument can be described by
its Jones matrix $\mathbf{J}_{\mbox{\tiny tot}}$, a $2\times 2$
complex matrix, which relates the radiation $\mathbf{e}_{\mbox{\tiny
    det}}$ that hits a detector to the incoming radiation on the
telescope $\mathbf{e}_{\mbox{\tiny sky}}$:
\begin{eqnarray}
\label{eq:jones_any}
	\mathbf{e}_{\mbox{\tiny det}}(\nu,\mathbf{n}) & = & \mathbf{J}_{\mbox{\tiny
            tot}}(\nu,\mathbf{n})\, \mathbf{e}_{\mbox{\tiny
            sky}}(\nu,\mathbf{n}) \nonumber \\
        & = & 
	\mathbf{J}_{\mbox{\tiny det}}\,
	\mathbf{J}_{\mbox{\tiny filter}}(\nu)\,  
        \mathbf{J}_{\mbox{\tiny beam}}(\nu,\mathbf{n})\,
        \mathbf{e}_{\mbox{\tiny sky}}
\end{eqnarray}
where $\nu$ is the electromagnetic frequency, $\mathbf{n}$ is the
direction of observation and we have decomposed the Jones matrix into
optical element Jones matrices. As the detector is sensitive to
polarization, we can write the associated Jones matrix
$\mathbf{J}_{\mbox{\tiny det}}$ as:
\begin{equation}
	\mathbf{J}_{\mbox{\tiny det}} = 
	\left(
		\begin{array}{cc} 1 & 0 \\ 0 & \sqrt{\eta} \end{array}
	\right),
\end{equation}
where $\eta$ is the cross-polarization leakage. We assume it
is independent of the frequency of the incoming radiation, which is
reasonable as it is mainly due to absorption of the cross-polarization
component on the edge of the absorbing grid \citep{Jones:2003mz}.

The filter can also be described by a Jones matrix, as it is not a
depolarizing element in the sense defined by \cite{Ditchburn:1976}. It
is simply given by:
\begin{equation}
\mathbf{J}_{\mbox{\tiny filter}} = \left(\begin{array}{cc}
  \sqrt{\tau(\nu)} & 0 \\ 0 & \sqrt{\tau(\nu)}\end{array}\right)
\end{equation}
where $\tau(\nu)$ is the bandpass transmission of the filter, which
has been measured accurately on ground.

Finally, the beam of both the telescope and the horns is described by
a generic Jones matrix, $\mathbf{J}_{\mbox{\tiny beam}}(\nu,
\mathbf{n})$, which depends on both radiation frequency and direction
on the sky. The electric field received by the detector is thus given
by:
\begin{equation}
\mathbf{e_{\mbox{\tiny det}}}(\nu,\mathbf{n}) =  \sqrt{\tau(\nu)}\,
\mathbf{R}\,\left(\begin{array}{cc} J_{xx} & J_{xy}
  \\ \sqrt{\eta}J_{yx} & \sqrt{\eta}J_{yy}
\end{array}\right)  \mathbf{R}^{-1}\mathbf{e_{\mbox{\tiny sky}}} (\nu,\mathbf{n})
\end{equation}
where we have included the matrix $\mathbf{R}$ which rotates the
incoming radiation from the sky reference frame to the intrument
reference frame. The coefficients $J_{ij}$, with $i$, $j$ in $\{x,y\}$,
are the elements of the beam Jones matrix $\mathbf{J}_{\mbox{\tiny
    beam}}(\nu,\mathbf{n})$.

The intensity measured by the detector is the sum of the intensities
coming from each direction and for each frequency:
\begin{equation}
I_{\mbox{\tiny det}} = \int\!\!\!\int\langle\mathbf{e_{\mbox{\tiny det}}}(\nu,\mathbf{n})^\dag \cdot
\mathbf{e}_{\mbox{\tiny det}}(\nu,\mathbf{n}) \rangle\, d\mathbf{n}\, d\nu.
\end{equation}
To describe the sky signal, we use the Stokes parameters $I$, $Q$, $U$
and $V$ \citep[see, \textit{e.g.},][]{Born:1964}:
\begin{equation}
\begin{split}
I(\nu,\mathbf{n}) & = \langle E_xE_x^*\rangle + \langle E_yE_y^*\rangle \\ 
Q(\nu,\mathbf{n}) & = \langle E_xE_x^*\rangle - \langle E_yE_y^*\rangle \\ 
U(\nu,\mathbf{n}) & = \langle E_xE_y^*\rangle + \langle E_yE_x^*\rangle \\ 
V(\nu,\mathbf{n}) & = -i\left( \langle E_xE_y^*\rangle - \langle E_yE_x^*\rangle\right).
\end{split}
\end{equation}
where $I$ is the intensity, $Q$ and $U$ charaterize the linear
polarization and $V$ the circular polarization of the sky
radiation. We define analogously the beam Stokes parameters as:
\begin{equation}
\begin{split}
\tilde{I}_\alpha(\nu,\mathbf{n})  & =  J_{\alpha x}J_{\alpha x}^* + J_{\alpha y}J_{\alpha y}^*\\
\tilde{Q}_\alpha(\nu,\mathbf{n})  & = J_{\alpha x}J_{\alpha x}^* - J_{\alpha y}J_{\alpha y}^*\\
\tilde{U}_\alpha(\nu,\mathbf{n})  & =  J_{\alpha x}J_{\alpha y}^* + J_{\alpha y}J_{\alpha x}^*\\
\tilde{V}_\alpha(\nu,\mathbf{n})  & =  -i\left(J_{\alpha x}J_{\alpha y}^* - J_{\alpha y}J_{\alpha x}^*\right)
\end{split}
\end{equation}
($\alpha = x,y$). Note that in general the beam Stokes parameters
depend on both frequency and sky direction. Therefore, we can write the
intensity measured by the detector as:
\begin{eqnarray}
\label{eq:photo}
I_{\mbox{\tiny det}} & = & \frac{1}{2}\int\!\!\!\int \tau(\nu) \left[
  I(\tilde{I}_x+\eta\tilde{I}_y) \right.\nonumber\\
& & {}+ Q \left[(\tilde{Q}_x+\eta\tilde{Q}_y)\cos 2\theta 
    + (\tilde{U}_x+\eta\tilde{U}_y)\sin 2\theta \right] \nonumber\\
& & {}+ U \left[- (\tilde{Q}_x+\eta\tilde{Q}_y)\sin 2\theta 
    + (\tilde{U}_x+\eta\tilde{U}_y)\cos 2\theta \right] \nonumber\\
& & \left.{}- V (\tilde{V}_x+\eta\tilde{V}_y) \right]
\end{eqnarray}
where $\theta$ is the angle of orientation between the sky and
detector reference frames, and we have not explicitly written the
dependency of radiation and beam Stokes parameters to frequency $\nu$
and direction $\mathbf{n}$ for clarity.

\section{Systematics for polarization \label{sec:syste}}

In Eq.~(\ref{eq:photo}) each term that couples to one of the
Stokes parameters may depend on the direction of observation,
$\mathbf{n}$, and on frequency in non trivial ways. Several other
instrumental effects could be added to give an accurate description
of a detector measurement, such as its time constant, noise or
pointing errors.

The final calibration and analysis of HFI data needs to address all
these effects and will rely on both ground and in-flight
measurements. This is beyond the scope of this paper. However, some
comments can already be made.

HFI beam patterns have been simulated with GRASP
\citep[see][]{Maffei:2010ly,Tauber:2010ay} and these simulations have
been verified by ground calibration performed by Thales Industries. It
was shown that \emph{optical} cross-polarization and circular
polarization $\tilde{V}$ due to telescope were less than 0.1\%. Their
impact has been studied separately \citep{Rosset:2007}.

We will thus consider in the following an ideal optical system for
which $\mathbf{J}_{\mbox{\tiny beam}}$ is proportional to the identity
matrix resulting in $\tilde{I}_x=\tilde{I}_y=\tilde{Q}_x=-\tilde{Q}_y$
and $\tilde{U}_x=\tilde{U}_y=\tilde{V}_x=\tilde{V}_y=0$.
Equation~\ref{eq:photo} therefore simplifies to
\begin{equation}
	I_{\mbox{\tiny det}}  =  \frac{1}{2} \iint \tau(\nu) \tilde{I}_x [(1+\eta)I+(1-\eta)(Q\cos 2\theta+U\sin 2\theta)] \, d\Omega d\nu.
\label{eq:photo_1}
\end{equation}

Realistic bandpasses and frequency dependence of optical beam coupling
terms are non-trivial effects that affect absolute calibration. More
specifically, calibration could depend on the electromagnetic spectrum
of the source. This is expected to impact component separation. In
this work, we focus on systematic effects on CMB polarization and rely
on absolute calibration on the CMB dipole, the amplitude f which is
known to 0.5\% accuracy \citep{Fixsen:1994}.  We expect to measure in
flight the relative gain to an accuracy of better than 0.2\%, given
the gain stability expected for HFI \citep[\textit{i.e.} better than
  WMAP, see][]{Hinshaw:2009os}. Beam asymmetries and pointing
errors couple to the scanning strategy of the instrument. A general
framework to assess their impact is presented in
\cite{Shimon:hz} and \cite{ODea:2007kx}.

Leaving these effects aside for this work, the measurement of a
detector reads:
\begin{equation}
	m = g\Big(I + \rho[ Q\cos2(\psi+\alpha)+U\sin2(\psi+\alpha)]\Big) + n
\label{eq:modelmeas}
\end{equation}
in which $n$ is the noise, $g$ is the total {\it gain}, $\rho =
(1-\eta)/(1+\eta)$ is commonly
referred to as \emph{polarization efficiency}, $\psi$ is the
dependence on the focal plane orientation on the sky and $\alpha$
stands for the relative \emph{detector orientation} with respect to it. 

\section{Propagation of errors for polarization calibration}
\label{sec:method}

In this section, we propagate errors on gain $g$, on polarization
efficiency $\rho$ and detector orientation $\alpha$ (as defined in the
previous section) up to Stokes parameters
(Sect.~\ref{sec:error_stokes}) and angular power spectra
(Sect.~\ref{sec:error_power_spectra}).

This method applies to all polarization experiments observing with
total power detectors such as HFI bolometers. It is close to the
approach taken by \cite{Shimon:hz} and \cite{ODea:2007kx}. A similar
approach, focused on coherent receivers, was first proposed by
\cite{Hu:2003yq}. The main difference of the method presented here is
that it addresses the specific case of Planck which combines
different detectors to determine $Q$ and $U$.

\subsection{Error on Stokes parameters \label{sec:error_stokes}}

Given a pixelization of the sky and gathering all samples that fall
into the same pixel $p$ in a vector $\mathbf{m}$, Eq.~\ref{eq:modelmeas}
generalizes to the usual matrix form:
\begin{equation}
\label{eq:m}
	m_t = A_{tp}s_p + n_t,
\end{equation}
in which $\mathbf{s} = (I,Q,U)$ is the pixelized polarized sky signal
and $\mathbf{n}$ represents the noise vector. The pointing matrix
$\mathbf{A}$ encodes both the direction of observation and the
photometric equation including the calibration parameters $g$, $\rho$
and $\alpha$. Projection of time-ordered data into a pixelized map is
done by solving Eq.~(\ref{eq:m}) for $\mathbf{s}$, knowing
$\mathbf{m}$ and the noise covariance matrix $N\equiv\langle
\mathbf{n} \mathbf{n}^T\rangle$. The maximum likelihood solution is
$\widehat{\mathbf{s}} = (\mathbf{A}^T \mathbf{N}^{-1}
\mathbf{A})^{-1}\mathbf{A}^T \mathbf{N}^{-1}\mathbf{m} = (\mathbf{A}^T
\mathbf{A})^{-1} \mathbf{A}^T\mathbf{m}$ if we consider only Gaussian,
white and piece wise stationary noise, as we shall do in the remaining
part of this work in order to focus on systematic effects.

We use a perturbative approach of assumed parameters $\tilde g$,
$\tilde\rho$ and $\tilde\alpha$ (leading to a pointing matrix
$\tilde{\mathbf{A}}$) around their true values $g$, $\rho$ and
$\alpha$ (leading to $\mathbf{A}$).  From Eq.~(\ref{eq:modelmeas}),
we can see that for $Q$ and $U$ Stokes parameters, errors on the gain
and polarization efficiency are degenerate. In the following, we use
an effective polarization efficiency $\rho^\prime \equiv g\rho$ and
keep $g$ for intensity only. The actual gain, polarization efficiency and
orientation for a given detector $d$ are therefore $g_d = \tilde g_d +
\gamma_d$, $\rho^\prime_d = \tilde{\rho_d}^\prime+\epsilon_d$ and
$\alpha_d = \tilde{\alpha_d} + \omega_d$ respectively\footnote{In the
  following, when a relation holds for $\gamma$, $\epsilon$ or
  $\omega$, we simply write $e$}. Thus, ignoring noise,
\begin{eqnarray}
	\hat{\mathbf{s}} 
	& = & 
	(\tilde{\mathbf{A}}^T\tilde{\mathbf{A}})^{-1}\tilde{\mathbf{A}}^T\mathbf{m} \label{eq:s_solve}
	\\
	& = & 
	\left[ \sum_d \tilde{\mathbf{A}}_d^T\tilde{\mathbf{A}}_d \right]^{-1} 
	\left[ \sum_d \tilde{\mathbf{A}}_d^T\mathbf{A}_d\right]\mathbf{s} 
	\\
	& \equiv & 
	\left[ \sum_d \tilde{\mathbf{A}}_d^T\tilde{\mathbf{A}}_d \right]^{-1} 
	\left[
		\sum_d \mathbf{\Lambda}_d(\gamma_d, \epsilon_d, \omega_d)
	\right]{\bf s}.
\label{eq:s}
\end{eqnarray}
In this expression, $\mathbf{\Lambda}_d(\gamma, \epsilon_d, \omega_d)$
is an explicit function of $\gamma$, $\epsilon_d$ and $\omega_d$, and
$\tilde g$, $\tilde{\rho}^\prime$ and $\tilde{\alpha}$ are only
parameters.

Considering small variations around $g$, $\rho^\prime$ and $\alpha$,
we can write the perturbative expansion to first order for both
$\gamma \ll 1$, $\epsilon \ll 1$ and $\omega \ll 1$:
\begin{eqnarray}
	\Delta \mathbf{s} 
		& = & \hat{\mathbf{s}} - \mathbf{s} \nonumber  \\
		& = & 
			\left[ \sum_d \tilde{\mathbf{A}}_d^T \tilde{\mathbf{A}}_d \right]^{-1}
			\left[ \sum_d \mathbf{\Lambda}_d(\gamma_d, \epsilon_d, \omega_d) - \mathbf{\Lambda}_d(0, 0, 0) \right] \mathbf{s} \nonumber \\
		& \simeq &
			\left[ \sum_d \tilde{\mathbf{A}}_d^T \tilde{\mathbf{A}}_d\right]^{-1}
			\sum_d \left[
				\frac{\partial \mathbf{\Lambda}_d}{\partial\gamma_d}\gamma_d +
				\frac{\partial
                                  \mathbf{\Lambda}_d}{\partial\epsilon_d}\epsilon_d
                                 \right.\nonumber\\
& & \qquad\qquad\qquad\qquad\qquad\qquad\qquad \left.
		{}+\frac{\partial \mathbf{\Lambda}_d}{\partial\omega_d}\omega_d
			\right] {\mathbf s}.
\label{eq:delta_s}
\end{eqnarray}
Partial derivatives with respect to gain $\gamma_d$, polarization
efficiency $\epsilon_d$ and orientation $\omega_d$ uncertainties are
derived in appendix \ref{se:appendix_deriv}.

The errors $\Delta\mathbf{s}$ strongly depend on the scanning strategy
through the number of hits per pixel and the distribution of detector
orientations. These are accounted for exactly by taking the scanning
strategy of the instrument and the positions of all detectors, and
computing the pointing-related quantities per pixel on which
$\mathbf{\Lambda}_d$ and its derivatives depend. This part of the work
may be intensive in terms of memory or disk access requirements
depending on which experiment is being modeled but needs to be
performed only once. Then, given a sky model, the generation of an
arbitrary large set of error maps $\Delta \mathbf{s}$ requires fewer resources and
involves only distributions of $\gamma_d$, $\epsilon_d$ and $\omega_d$.

Note that in the particular case of an experiment whose scanning
strategy is such that each detector observes each pixel of the map
under angles uniformly distributed over $[0,\pi]$, making a combined
map as in Eq.~(\ref{eq:s_solve}) is equivalent to making one set of
$I$, $Q$, and $U$ maps per detector and co-adding them to obtain the
final optimal maps of the experiment. In that case, sums of cosines
and sines vanish, which means that off diagonal terms of
$\tilde{\mathbf{A}}^T_d\tilde{\mathbf{A}}_d$ are zero and
Eq.~(\ref{eq:delta_s}) reads simply
\begin{equation}
	\Delta \mathbf{s} = 
	\langle \gamma{\rangle}_d \left( \begin{array}{c} I\\0\\0\end{array} \right) + 
	\langle \frac{\epsilon}{\rho^\prime}{\rangle}_d \left( \begin{array}{c} 0\\Q\\U\end{array} \right) + 
	2\langle \omega{\rangle}_d \left( \begin{array}{c}0\\U\\-Q\end{array} \right)
\label{eq:map_error_ideal}
\end{equation}

Because of the linearization, the final map is sensitive to the
averages of these parameters.  If errors are correlated (or identical
at worst), they do not average down; if they are randomly distributed around zero
mean, they do. These results are in agreement with
\cite{ODea:2007kx}. As we will see in the Sect.~\ref{sec:hfi}, this
is not the case for HFI, for which none of these simplifications
applies.

\subsection{Errors on Angular Power Spectra \label{sec:error_power_spectra}}

Following conventions of \cite{Zaldarriaga:97}, the projection onto
spherical harmonics of intensity and polarization reads:
\begin{eqnarray}
  a_{\ell m}^T & = & \int I(\mathbf{n}) Y^*_{lm}(\mathbf{n})\;d\mathbf{n}, \nonumber \\
  a_{\ell m}^E & = & - \int \left[Q(\mathbf{n})R^+_{lm}(\mathbf{n}) + iU(\mathbf{n})R^-_{lm}(\mathbf{n})\right]\;d\mathbf{n},\nonumber \\
  a_{\ell m}^B & = & i \int \left[Q(\mathbf{n})R^-_{\ell m}(\mathbf{n}) + iU(\mathbf{n})R^+_{lm}(\mathbf{n})\right]\;d\mathbf{n} \nonumber 
\end{eqnarray}
where the $R^{\pm}_{lm} = \ _{2}Y^*_{lm} \pm \ _{-2}Y^*_{lm}$ depend on
the $s$-spin spherical harmonic functions $\ _{s}Y_{lm}(\mathbf{n})$
($s=\{0,2,-2\}$).

Spherical harmonics transforms are linear, so derivatives of $\mathbf{a}_{\ell m} =
\left( a^T_{\ell m}, a^E_{\ell m}, a^B_{\ell m} \right)$ read
\begin{eqnarray}
	\frac{\partial \mathbf{a}_{\ell m}}{\partial e}
	& = &
	\int \left( \begin{array}{ccc}
		Y^*_{lm} & 0 & 0 \\
		0 & -R^+_{lm} & - i R^-_{lm} \\
		0 & i R^-_{lm} & - R^+_{\ell m}
	\end{array} \right)_{(\mathbf{n})}
  	\frac{\partial \mathbf{s}}{\partial e}(\mathbf{n}) \; d\mathbf{n} \nonumber
\label{aeq:dalm}
\end{eqnarray}

We use a simple pseudo-$C_\ell$ estimator, $\tilde{C}_\ell$, which is
$\chi^2$-distributed with a mean equal to the underlying $C_\ell$,
$\nu_\ell = (2\ell+1)$ degrees of freedom and a variance of $2
C_\ell / \nu_\ell$:
\begin{equation}
	\tilde{C}^{XY}_{\ell} = \frac{1}{(2\ell+1)}\sum_{m=-\ell}^{\ell}a^{X*}_{\ell m}a^Y_{\ell m}.
\end{equation}

This estimator neglects the E-B mixing due to incomplete sky coverage
\citep{Lewis:02} and assumes a cross-power spectrum for which noise bias
is null (or if auto-spectra are used, that the noise bias has been
previously removed) because their interaction with the systematic
effects introduced here are of second order.

Using the previous relations, straightforward algebra leads from Eq.~(\ref{eq:delta_s}) to its counterpart in harmonic space:
\begin{eqnarray}
	&&\Delta \tilde{C}_\ell  
	=
		\sum_d \frac{\partial \tilde{C}_\ell}{\partial\gamma_d} \gamma_d + 
		\sum_d \frac{\partial \tilde{C}_\ell}{\partial\epsilon_d} \epsilon_d + 
		\sum_d \frac{\partial \tilde{C}_\ell}{\partial\omega_d} \omega_d \nonumber\\
	&& {}+
		\frac{1}{2} \sum_{d,d'} \left[
			\frac{\partial^2 \tilde{C}_\ell}{\partial\gamma_d\partial\gamma_{d'}} \gamma_d\gamma_{d'} + 
			\frac{\partial^2 \tilde{C}_\ell}{\partial\epsilon_d\partial\epsilon_{d'}} \epsilon_d\epsilon_{d'} \right.\nonumber\\
&& \quad\qquad\qquad\qquad\qquad\qquad\qquad \left.
	{}+\frac{\partial^2 \tilde{C}_\ell}{\partial\omega_d\partial\omega_{d'}} \omega_d\omega_{d'}
		\right],
\label{eq:dcl}
\end{eqnarray}
where, for $e = \gamma$, $\epsilon$ or $\omega$,
\begin{eqnarray}
	\frac{\partial C^{XY}_\ell}{\partial e} 
	& = &
	\frac{1}{2\ell+1}\sum_{m=-\ell}^{\ell}
	\left[
		\frac{\partial a^{X*}_{\ell m}}{\partial e}a^Y_{\ell m} + a^{X*}_{\ell m}\frac{\partial a^Y_{\ell m}}{\partial e}
	\right]
	\label{eq:dcldepsilon}\\
	\frac{\partial^2 C^{XY}_\ell}{\partial e \partial e'} 
	& = &
	\frac{1}{2\ell+1} \sum_{m=-\ell}^{\ell} \left[\frac{\partial^2a^{X*}_{\ell m}}{\partial e\partial e'} a^Y_{\ell m} +
	\frac{\partial a^{X*}_{\ell m}}{\partial e}\frac{\partial a^{Y}_{\ell m}}{\partial e'} \right.\nonumber \\
	& & \quad\quad\quad\quad\quad\left.{}+ 
	\frac{\partial a^{X*}_{\ell m}}{\partial e'}\frac{\partial a^{Y}_{\ell m}}{\partial e} + 
	a^{X*}_{\ell m}\frac{\partial^2a^Y_{\ell m}}{\partial e\partial e'}\right].
	\label{eq:d2cldepsilon2}
\end{eqnarray}

We ignore cross-terms between different systematic parameters so the
previous expressions are only applicable when all but one of the
parameters are set to zero. The cross-terms have been checked to be
one order of magnitude below the direct terms. Note that we push the
perturbative expansion to second order, since $E$-modes are much larger
than $B$-modes and a second order effect on $E$-modes has an impact
comparable to a first order effect on $B$-modes.

\subsection{Monte-Carlo Simulations}

We have now everything in hand to perform the semi-analytical estimate of the polarization calibration systematic effects.
The method can be described in 5 main steps:
\begin{enumerate}
	\item From the scanning strategy of the instrument, for each detector $d$, project into a map : $\cos 2\psi$, $\sin 2\psi$, $\cos 2\psi \sin 2\psi$, and $\cos^2 2\psi$.
	\item With these quantities, compute for each pixel of the map the following $3\times3$ matrices: $\left[\sum_d \tilde{\mathbf{A}}_d^T\tilde{\mathbf{A}}_d\right]^{-1}$, $\mathbf{\Lambda}_d$, and its first and second derivatives.
	\item Use a simulated CMB sky $\mathbf{s}$ and Eq.~(\ref{eq:delta_s}) to compute partial derivatives $\partial \mathbf{s} / \partial e$ (up to second order)
	\item Compute all cross-power spectra between $\mathbf{s}$ and its derivatives
	\item Combine these results using gaussian random distributions of $\gamma_d$, $\epsilon_d$ and $\omega_d$ (with various rms $\sigma$) in Eq.~(\ref{eq:dcl}) to obtain the final error on the angular power spectrum.
\end{enumerate}

The power spectra estimator used is a pseudo-$C_\ell$ estimator based
on the cross-power spectra algorithm \citep{Tristram:2005ij}, extended
to polarization \citep{kogut:03,grain:09}. The semi-analytic method
described in this section has been compared to full Monte-Carlo
simulations and gives results compatible with statistical expectations
for the number of simulations performed.

\section{Application to Planck-HFI focal plane}
\label{sec:hfi}

We apply the method described in the previous section to the Planck
HFI to set requirements on gain, polarization efficiency and
orientation. We simulated HEALPix \citep{Gorski:2005rt} full-sky maps
at a resolution of $\sim3.5$~arcmin ($nside=1024$) so that all pixels
are seen and each pixel is uniformly sampled. This avoids the
complications of estimating power spectra on a cut sky when allowing
for the same conclusions, as our power spectrum estimator is not
biased in the mean. The scanning strategy that we use is a realistic
simulation of what Planck will actually do in a 14-month mission. The
sky signal is pure CMB simulated from the best $\Lambda CDM$ fit to
WMAP 5 years data \citep{Dunkley:2009bz} with $r=0.05$, supposing the
CMB signal to be dominant over foregrounds residuals (at least for
intensity and $E$-mode CMB signals).

As described in Sect.~\ref{sec:detfp}, the Planck scanning strategy
and focal plane design do not allow the data from a single PSB pair to
provide independent maps of the Stokes parameters.  Here, we will use
two PSB pairs calibrated in intensity and consider small variations
around their gain $g_d = 1$, nominal angles $\alpha_d = \{0\degres,
90\degres, 45\degres, 135\degres\}$ and nominal polarization
efficiency $\rho^\prime_d = 1$ (corresponding to perfect PSB).

\subsection{Error on Stokes parameter for HFI \label{sec:hfi_stokes_error}}

We refer to appendix~\ref{se:appendix_deriv} for the explicit form of
the derivative terms of the Stokes parameters. Here, we emphasize the
issues specific to \hbox{HFI}. In this case,
Eq.~(\ref{eq:delta_s}) reads (see
Eqs.~\ref{aeq:dLdr}-\ref{aeq:d2Ldrde})
\begin{eqnarray} 
	\Delta \mathbf{s} 
	& = & 
	\left( \begin{array}{ccc}
		\Delta_{II} & \Delta_{IQ} & \Delta_{IU} \\ 
		\Delta_{QI} & \Delta_{QQ} & \Delta_{QU}\\
		\Delta_{UI} & \Delta_{UQ} & \Delta_{UU}
	\end{array} \right) \mathbf{s}.
\label{eq:mixing_matrix}
\end{eqnarray}

For gain variations only, non-zero elements of the matrix are given
for each pixel, to first order, by
\begin{eqnarray}
	\Delta_{II}^{g} & = & \textstyle
			\frac{1}{4} (\gamma_1 + \gamma_2 + \gamma_3 + \gamma_4)
	\label{eq:error_II_gain} \\
	\Delta_{QI}^{g} & = & \textstyle
			\frac{1}{4} (\gamma_1-\gamma_{2}) \left< \cos2\psi \right> - 
			\frac{1}{4} (\gamma_{3}-\gamma_{4}) \left< \sin 2\psi \right>
	\label{eq:error_QI_gain} \\
	\Delta_{UI}^{g} & = & \textstyle
			\frac{1}{4} (\gamma_{1}-\gamma_{2}) \left< \sin 2\psi \right> + 
			\frac{1}{4} (\gamma_{3}-\gamma_{4}) \left< \cos 2\psi \right>
	\label{eq:error_UI_gain}
\end{eqnarray}

For polarization efficiency only, elements of the matrix are given for
each pixel, to first order, by
\begin{eqnarray}
	\Delta_{IQ}^{\rho} & = & \textstyle
			\frac{1}{4} (\epsilon_1-\epsilon_{2}) \left< \cos2\psi \right> - 
			\frac{1}{4} (\epsilon_{3}-\epsilon_{4}) \left< \sin 2\psi \right>
	\label{eq:error_IQ_xpol} \\
	\Delta_{IU}^{\rho} & = & \textstyle
			\frac{1}{4} (\epsilon_{1}-\epsilon_{2}) \left< \sin 2\psi \right> + 
			\frac{1}{4} (\epsilon_{3}-\epsilon_{4}) \left< \cos 2\psi \right>
	\label{eq:error_IU_xpol} \\
	\Delta_{QQ}^{\rho} & = & \textstyle
			\frac{1}{2} (\epsilon_{1}+\epsilon_{2}) \left< \cos^2 2\psi \right> + 
			\frac{1}{2} (\epsilon_{3}+\epsilon_{4}) \left< \sin^2 2\psi \right>
	\label{eq:error_QQ_xpol} \\
	\Delta_{QU}^{\rho} & = & \textstyle 
			\frac{1}{2}
			\left[ (\epsilon_{1}+\epsilon_{2}) - (\epsilon_{3}+\epsilon_{4}) \right]
			\left< \cos 2\psi \sin 2\psi \right>
	\label{eq:error_QU_xpol} \\
	\Delta_{UQ}^{\rho} & = & \textstyle
			\frac{1}{2}
			\left[ (\epsilon_{1}+\epsilon_{2}) - (\epsilon_{3}+\epsilon_{4}) \right]
			\left< \cos 2\psi \sin 2\psi \right>
	\label{eq:error_UQ_xpol} \\
	\Delta_{UU}^{\rho} & = & \textstyle
			\frac{1}{2} (\epsilon_{1}+\epsilon_{2}) \left< \sin^2 2\psi \right> + 
			\frac{1}{2} (\epsilon_{3}+\epsilon_{4}) \left< \cos^2 2\psi \right>.
	\label{eq:error_UU_xpol}
\end{eqnarray}

In the case of orientation errors only, to first order,
\begin{eqnarray}
	\Delta_{IQ}^{\alpha} & = & \textstyle
			- \frac{1}{2} (\omega_{1}-\omega_{2}) \left<\sin 2\psi \right>
			- \frac{1}{2} (\omega_{3}-\omega_{4}) \left<\cos 2\psi \right>
	\label{eq:error_IQ_orientations} \\
	\Delta_{IU}^{\alpha} & = & \textstyle
			\frac{1}{2} (\omega_{1}-\omega_{2}) \left<\cos 2\psi \right> - 
			\frac{1}{2} (\omega_{3}-\omega_{4}) \left<\sin 2\psi \right>
	\label{eq:error_IU_orientations} \\
	\Delta_{QQ}^{\alpha} & = & \textstyle
			- 
			\left[ (\omega_{1}+\omega_{2}) - (\omega_{3}+\omega_{4}) \right]
			\left< \cos 2\psi \sin 2\psi \right>
	\label{eq:error_QQ_orientations} \\
	\Delta_{QU}^{\alpha} & = & \textstyle
			(\omega_{1}+\omega_{2}) \left< \cos^2 2\psi \right> + 
			(\omega_{3}+\omega_{4}) \left< \sin^2 2\psi \right>
	\label{eq:error_QU_orientations} \\
	\Delta_{UQ}^{\alpha} & = & \textstyle
			-(\omega_{1}+\omega_{2}) \left< \sin^2 2\psi \right>\!
			-(\omega_{3}+\omega_{4}) \left< \cos^2 2\psi \right>
	\label{eq:error_UQ_orientations} \\
	\Delta_{UU}^{\alpha} & = & \textstyle
			\left[ (\omega_{1}+\omega_{2}) - (\omega_{3}+\omega_{4}) \right]
			\left< \cos 2\psi \sin 2\psi \right>.
	\label{eq:error_UU_orientations}
\end{eqnarray}

\begin{figure}[!h]
	\begin{center}
	\includegraphics[height=120pt]{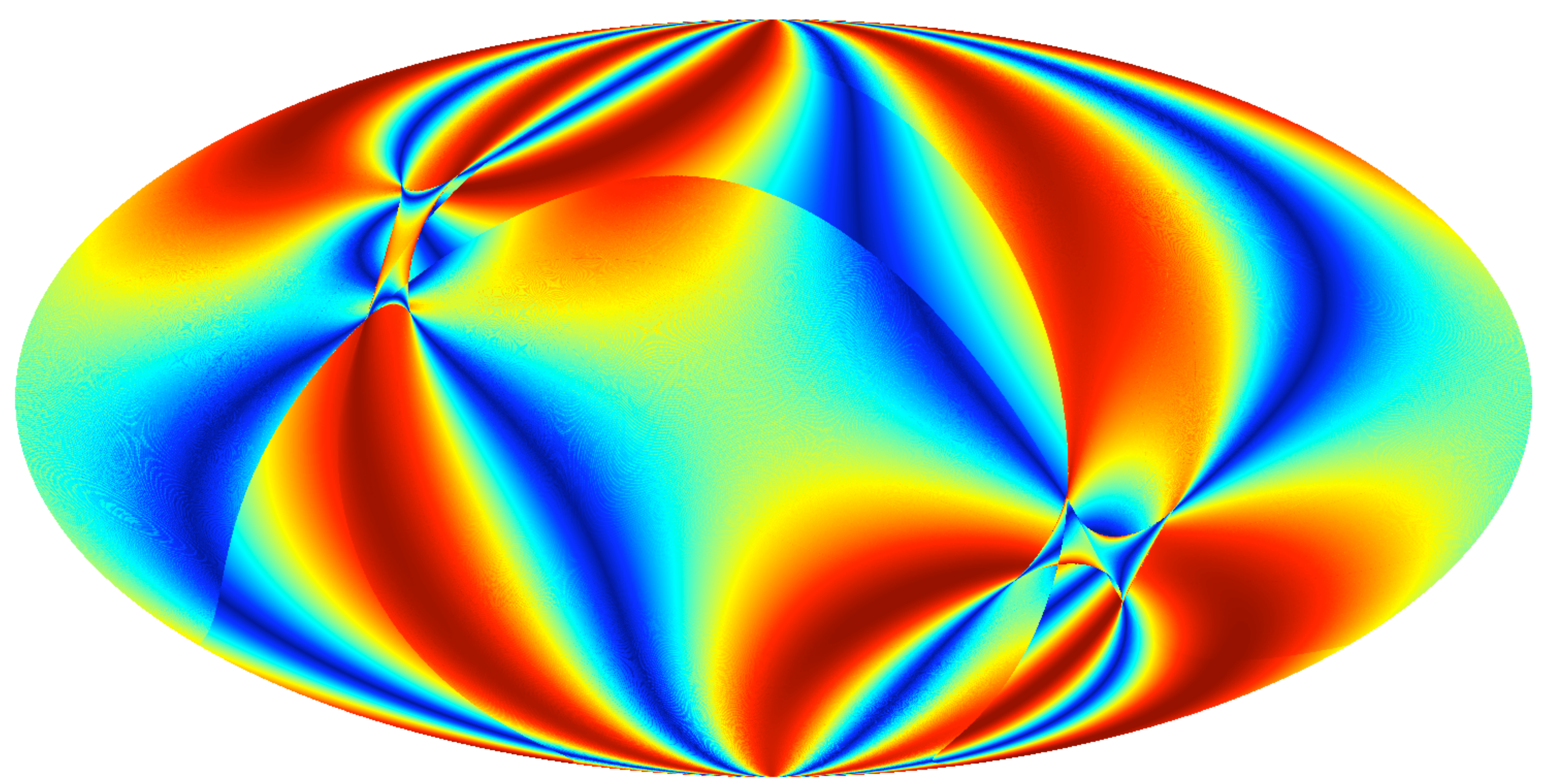}
	\includegraphics[height=120pt]{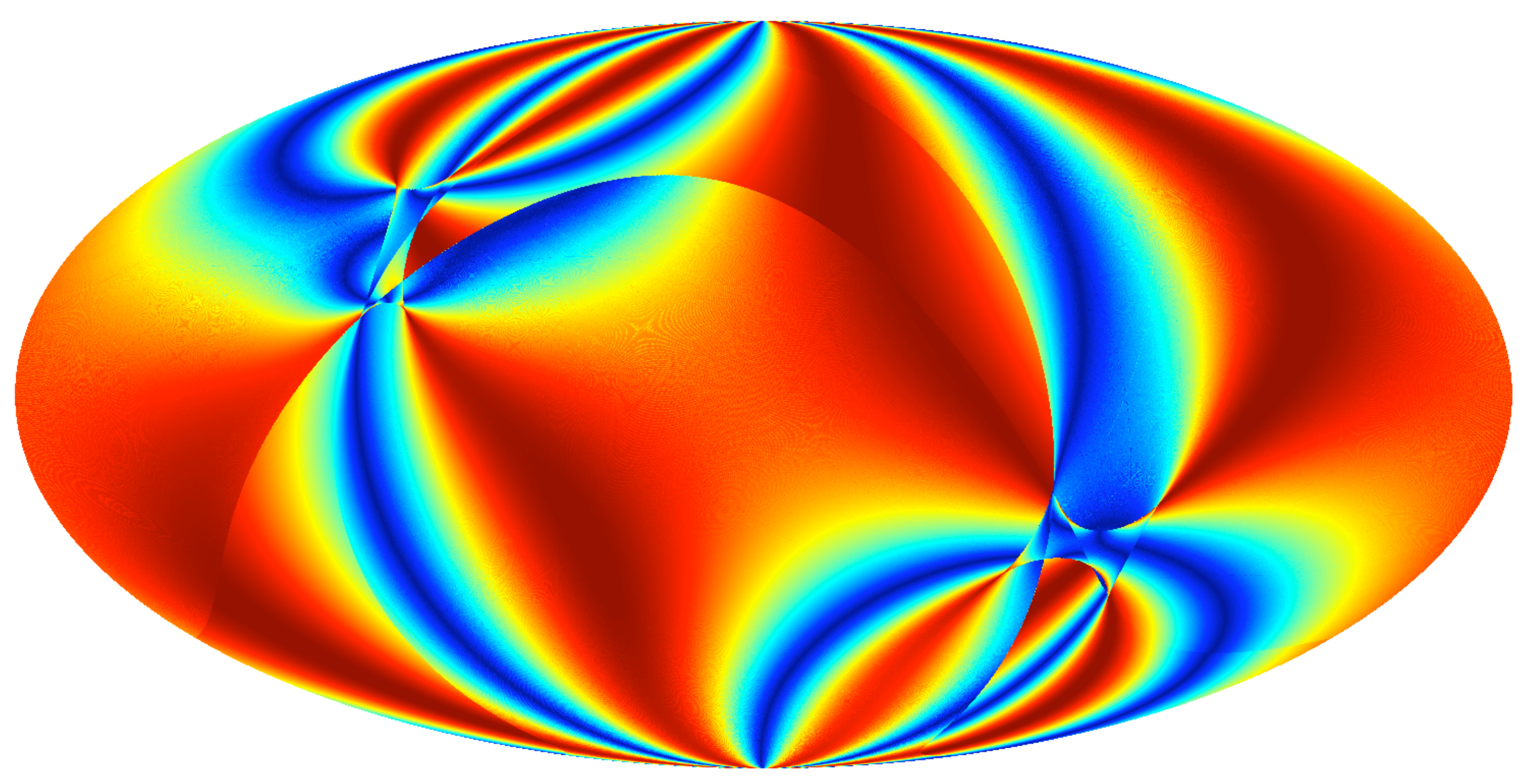}\\
	\includegraphics[height=120pt]{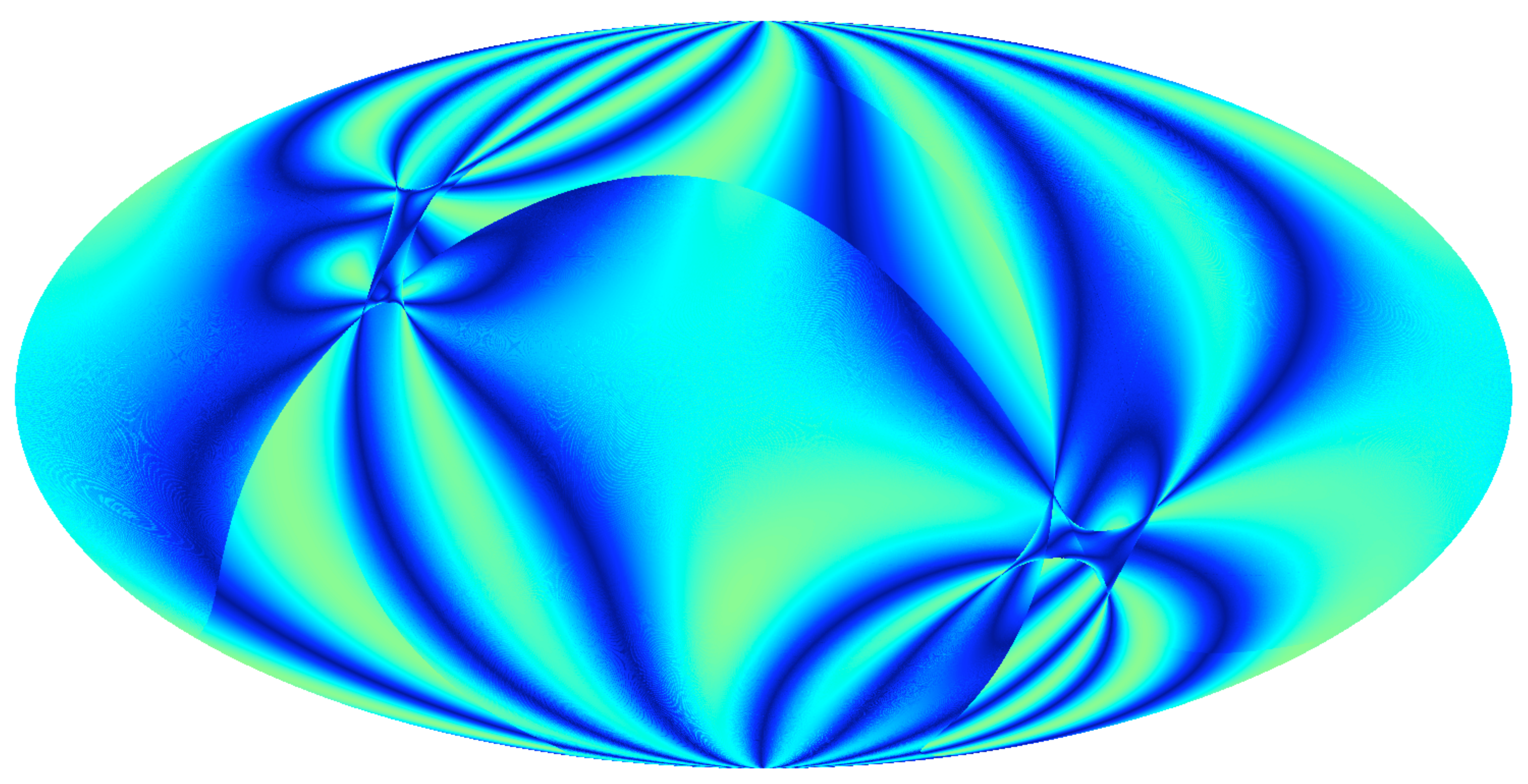}
	\includegraphics[height=120pt]{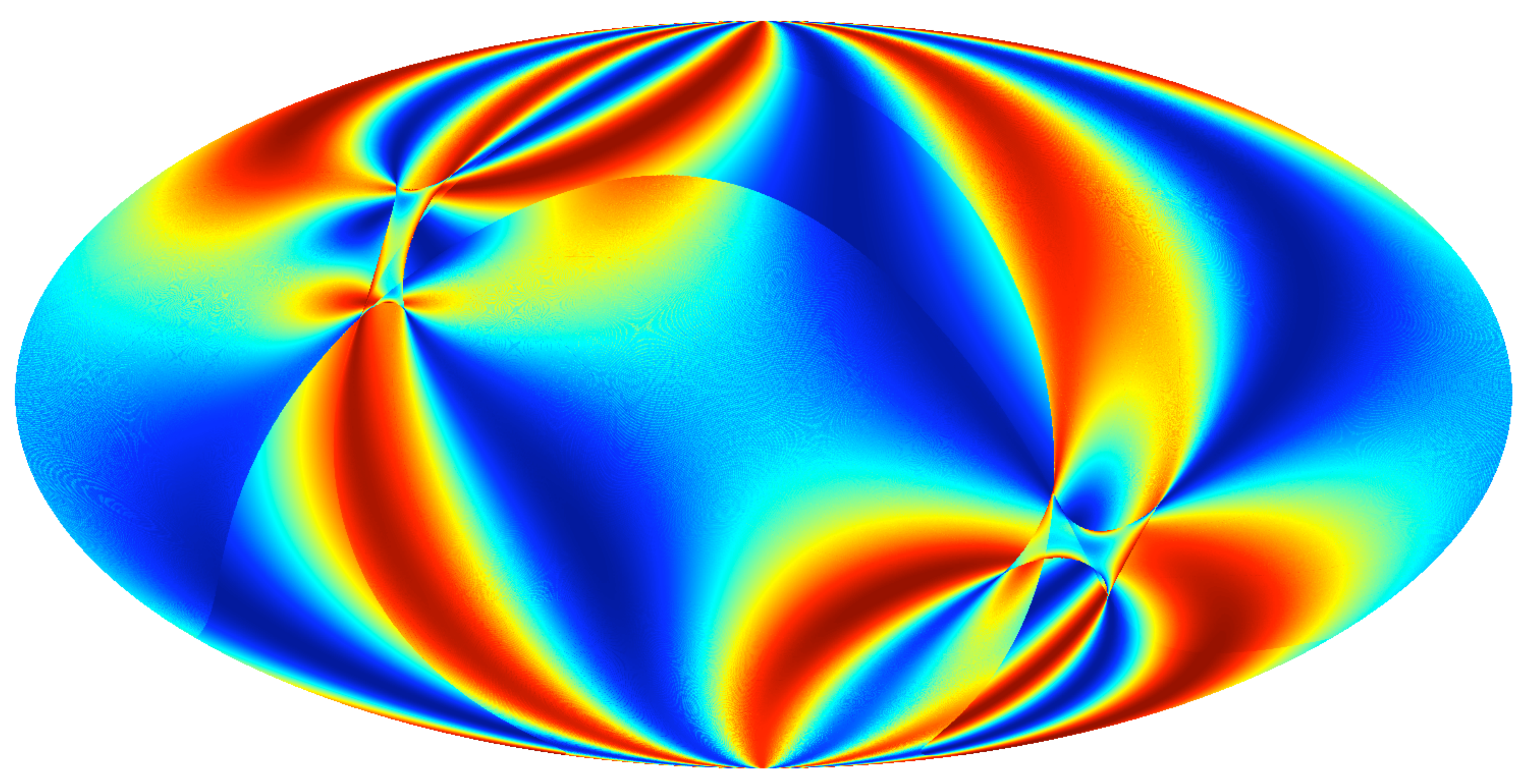}
	\includegraphics[width=6cm]{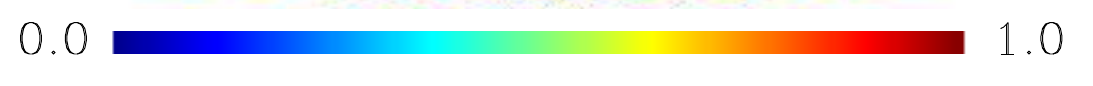}
	\caption{ \footnotesize Amplitude of the various terms in
          Eqs.~(\ref{eq:error_II_gain}-\ref{eq:error_UU_orientations})
          describing the focal plane angle distribution on the sky for
          a mock but realistic Planck scanning coverage (HEALPix maps
          at $nside=1024$, Galactic coordinates). {\it From top to bottom :} $|\langle \cos
          2\psi \rangle|$, $|\langle \sin 2\psi \rangle|$, $|\langle \sin
          4\psi \rangle|/2$, $\langle \cos^2 2\psi \rangle$.}
	\label{fig:angle_on_the_sky}
	\end{center}
\end{figure}

In these
Eqs~(\ref{eq:error_II_gain}-\ref{eq:error_UU_orientations}), the
average is over the samples falling into a given pixel. It depends
only on the scanning strategy. Figure~\ref{fig:angle_on_the_sky} shows
the angle distribution on the sky for a realistic Planck
scanning strategy.  Planck shows large inhomogeneities that induce
additional terms with respect to the case of a single bolometer.

\begin{description}

\item[{\bf Leakage from intensity to polarization.}] Error on gain only produces leakage from intensity to polarization (see eq~\ref{aeq:dLdg}).
This leakage is driven by the relative errors inside a given horn which indicates that an absolute error on the gain (same for all detectors) will not produce any leakage.
Neither polarization efficiency nor detector orientation errors induce any leakage from $I$ into polarization $Q$ and $U$ (see Eqs.~\ref{aeq:dLdr}-\ref{aeq:d2Ldrde}).\\

\item[{\bf Leakage from polarization to intensity.}] Both polarization
efficiency and orientation error produce leakage from polarization to
intensity. It is driven by the difference of errors within one horn
and the relative weight of each horn depends on the distribution of
$\psi$ (see Fig.~\ref{fig:angle_on_the_sky}).\\

\item[{\bf Polarization mixing.}] Polarization calibration parameters
  mix both $Q$ and $U$. This means that they induce leakage from $Q$
  to $U$ through the term $\Delta^\rho_{QU}$ (and from $U$ to $Q$
  through the term $\Delta^\rho_{UQ}$) but also alter the amplitude of
  polarization ($\Delta^\rho_{QQ}$ and $\Delta^\rho_{UU} \ne 0$). If
  we consider identical errors for each detector, we are in the
  limiting case where orientation error induces only leakage
  (Eqs.~\ref{eq:error_QU_orientations},\ref{eq:error_UQ_orientations})
  and polarization efficiency only changes the amplitude of
  polarization (Eqs.~\ref{eq:error_QQ_xpol},\ref{eq:error_UU_xpol}) as
  described by Eq.~\ref{eq:map_error_ideal}. In the case of
  Planck-HFI, and considering independent errors, none of these
  simplifications apply. In particular, different parameter averages
  from one horn to the other induce both $Q$ and $U$ mixing and
  amplitude modification.\\

\end{description}

\subsection{Results for $E$ and $B$-mode power spectra \label{sec:error_planck_ps}}

The semi-analytical method described in Sect.~\ref{sec:method} is
able to propagate instrumental errors up to the six CMB power spectra:
$TT$, $EE$, $BB$, $TE$, $TB$ and $EB$. In this section, we will focus
on the $E$ and $B$-mode power spectra and discuss results obtained for
Planck-HFI in case of absolute (Sect.~\ref{sec:ps_absolute_error}) and
relative uncertainties (Sect.~\ref{sec:ps_relative_error}).  Other
spectra (like $TB$ and $EB$) that are predicted to be null for CMB
signal, can be very useful in revealing ``leakage'' due to
systematics. However, many systematic effects can produce such
leakage, which will make their separate identification very
complicated when using only these modes.

\subsubsection{ Global error over the focal plane / calibration on the sky \label{sec:ps_absolute_error}}

Absolute calibration of total power is done using the orbital dipole
that has the same electromagnetic spectrum as the CMB and is not
degenerate with the underlying sky signal as its sign changes after 6
months of observation. From Eqs.~(\ref{eq:error_QI_gain})
and~(\ref{eq:error_UI_gain}), absolute error on the gain $g$ will not
produce any leakage in polarization signals:
\begin{eqnarray}
	\Delta^g \mathbf{s}
	& = & 
	\left( \begin{array}{ccc}
		\gamma & 0 & 0 \\
		0 & 0 & 0\\
		0 & 0 & 0
	\end{array} \right) \mathbf{s} \text{\hspace{0.5cm} for gain}.
\end{eqnarray}
As far as polarization is concerned, we need a polarized source on the
sky.  The Crab nebulae, a supernova remnant, is a good candidate as it
shows a large polarization emission in the Planck-HFI frequency
bands. It has been observed in a wide range of frequencies and shown
to have polarization properties stable enough to be a calibrator for
polarization experiments. Dedicated observations of this source were
done by IRAM at 89~GHz \citep{Aumont:2009rc}. The impact of an
approximate knowledge of the polarization sky calibrator leads to a
uniform error over the focal plane. In this case, the $\omega$ and
$\epsilon$ parameters do not depend on the detector. From
Eqs.~\ref{eq:error_IQ_xpol}-\ref{eq:error_UU_orientations}, we found
that the intensity does not leak into polarization with polarization
efficiency and orientation errors ($\Delta_{IQ} = \Delta_{IU} = 0$)
and
\begin{eqnarray}
	\Delta^\rho \mathbf{s}
	& = & 
	\left( \begin{array}{ccc}
		0 & 0 & 0 \\
		0 & \epsilon & 0\\
		0 & 0 & \epsilon
	\end{array} \right) \mathbf{s} \text{\hspace{0.5cm} for polarization efficiency},
\end{eqnarray}

\begin{eqnarray}
	\Delta^\alpha \mathbf{s} 
	& = & 
	\left( \begin{array}{ccc}
		0 & 0 & 0 \\ 
		0 & \cos2\omega & \sin2\omega\\
		0 & -\sin2\omega & \cos2\omega
	\end{array} \right) \mathbf{s} \text{\hspace{0.5cm} for orientation}.
\end{eqnarray}

In terms of power spectra, an error in polarization efficiencies only
affects the amplitude of the $E$ and $B$ power spectra but does not
result in leakage from $E$ to $B$. On the other hand, an error in
orientations mixes $Q$ and $U$ maps resulting in both a leakage from
$E$ into $B$ (as well as $B$ into $E$) and a modification of $E$ and
$B$ amplitudes. However, as the $E$-mode signal is far above that of
the $B$-mode in amplitude, $\Delta C_\ell$ is dominated by $E$-mode to
second order:
\begin{equation}
\Delta C^X_\ell = 2\epsilon C^X_\ell + 4\omega^2 C^E_\ell,
\end{equation}
for X either $E$ or $B$-mode.

Consequently, for $E$-mode, the polarization efficiency uncertainty
must be $\epsilon < 0.5\%$ and the orientation uncertainty $\omega <
2\pdeg9$ to obtain less than 1\% error on the power spectrum
amplitude. Alternatively, the leakage is kept under 10\% of the cosmic
variance if $\epsilon < 0.3\%$ and $\omega < 2\pdeg1$ for $\ell = 2 -
1000$.

To go further and target the $B$-mode signal, we show that the
orientation must be known to better than 1\pdeg3 (0\pdeg4) in order to
keep the leakage from $E$ to $B$-mode lower than 10\% (1\%) of the
expected $C_\ell^{B}$ for a tensor-to-scalar ratio of $r=0.05$ at
large angular scales ($\ell<100$). The error on its amplitude will be
driven by the polarization efficiency uncertainty ($2\epsilon$).

\subsubsection{ Relative Calibration between detectors \label{sec:ps_relative_error}}

As discussed in Sect.~\ref{sec:hfi_stokes_error}, there is no
generic case concerning the \emph{a priori} distribution of errors
for polarization parameters on HFI. We therefore performed $10 000$
Monte-Carlo simulations to propagate the errors through to the E
and B polarized angular power spectra. Errors were drawn from a gaussian
distribution with various dispersions $\sigma_\gamma$,
$\sigma_\epsilon$ and $\sigma_\omega$ per detector. We then propagated
those uncertainties through to the $E$ and $B$ angular power spectra.

\begin{figure}[!h]
	\begin{center}
	\includegraphics[width=9cm]{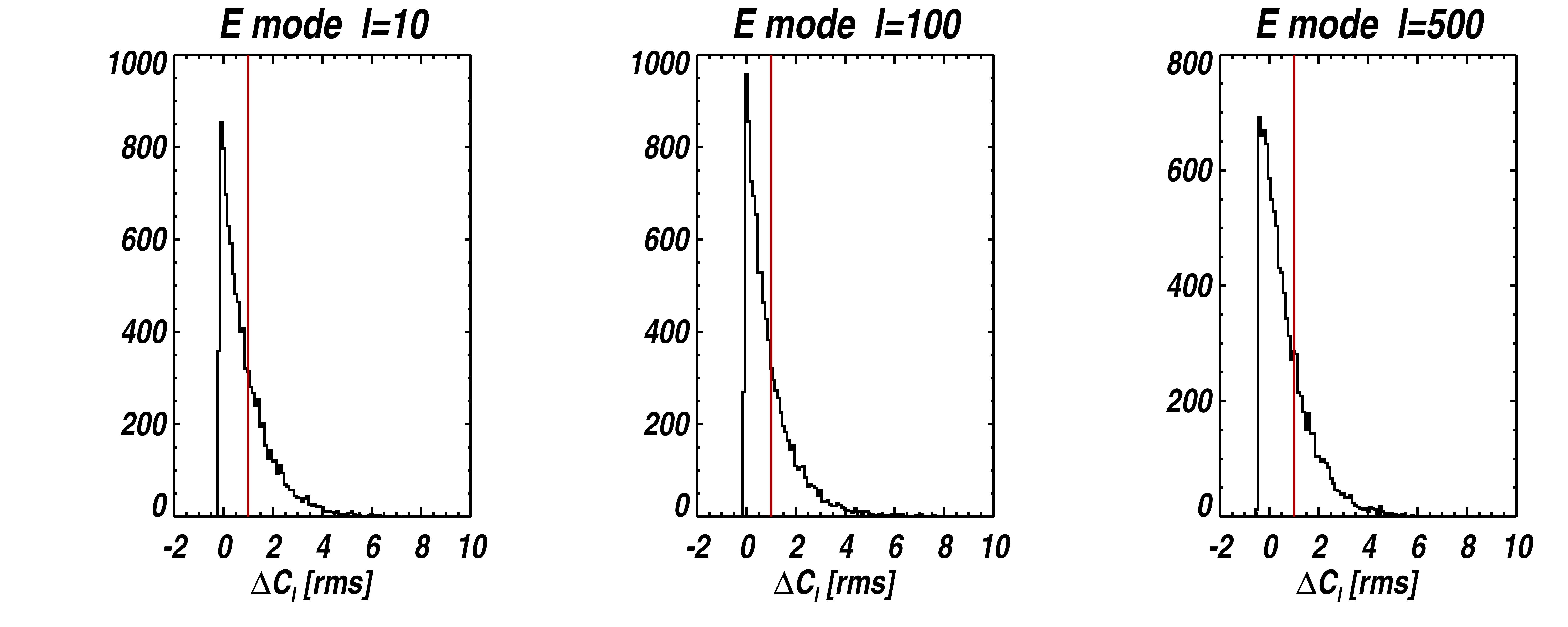}
	\includegraphics[width=9cm]{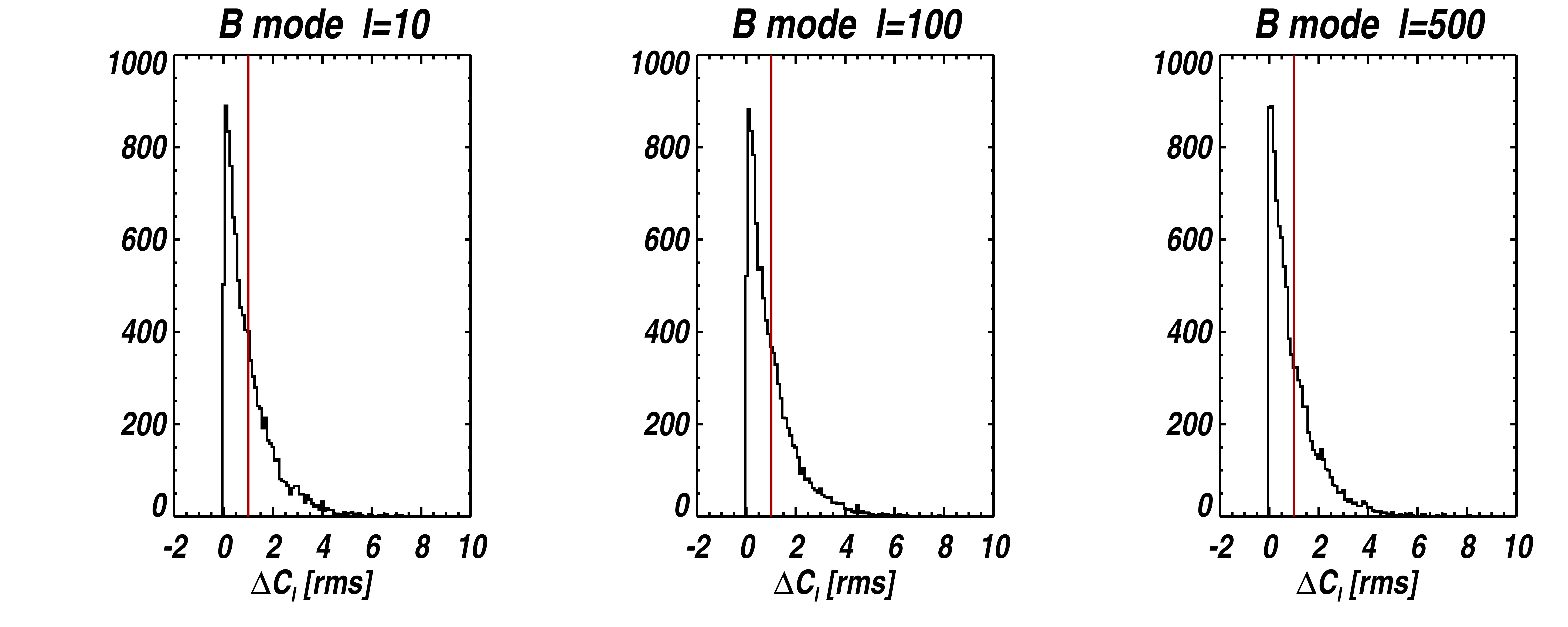}
	\caption{ \footnotesize Distribution of $\Delta C_\ell^{EE}$
          ({\it top}) and $\Delta C_\ell^{BB}$ ({\it bottom}) for
          $\sigma_\gamma = 0.2\%$ gain errors for multipoles
          $\ell=10$, $\ell=100$, $\ell=500$, normalized to their rms
          (\textit{red line}).}
	\label{fig:distrib_gain}
	\end{center}
\end{figure}

\begin{figure}[!h]
	\begin{center}
	\includegraphics[width=9cm]{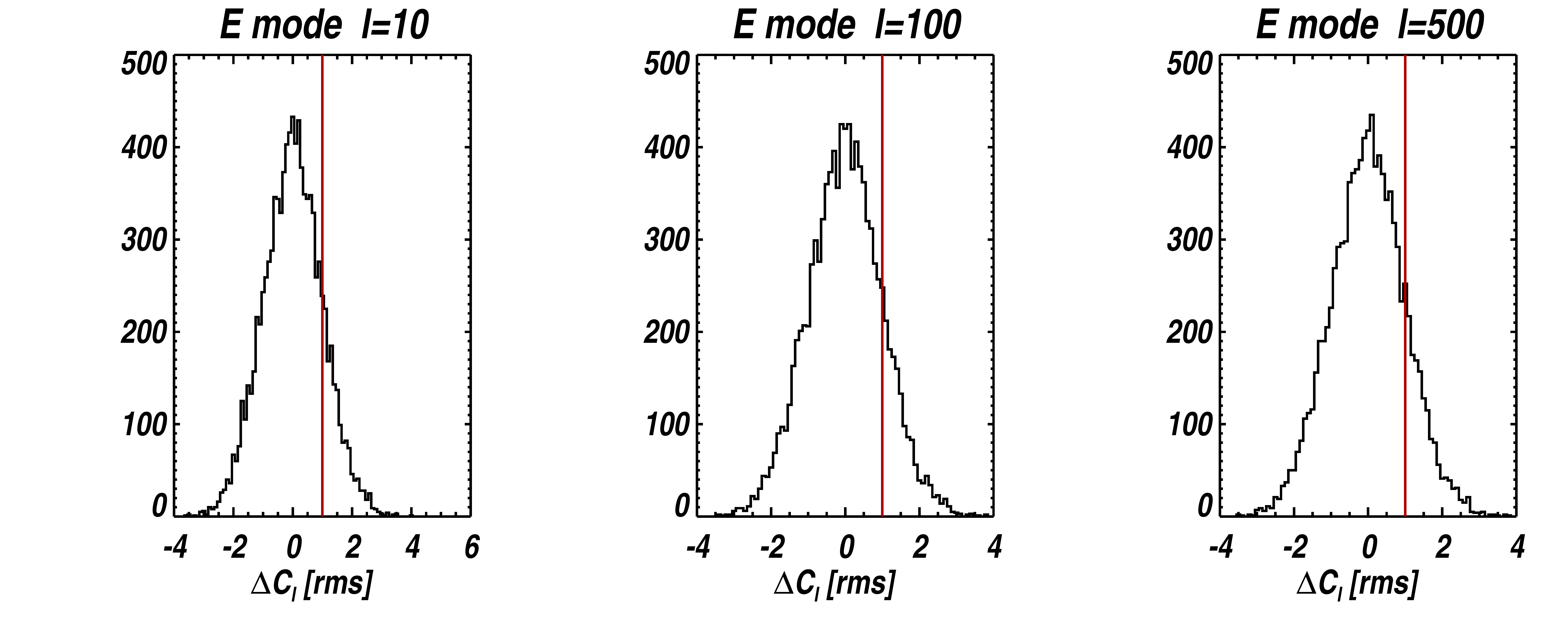}
	\includegraphics[width=9cm]{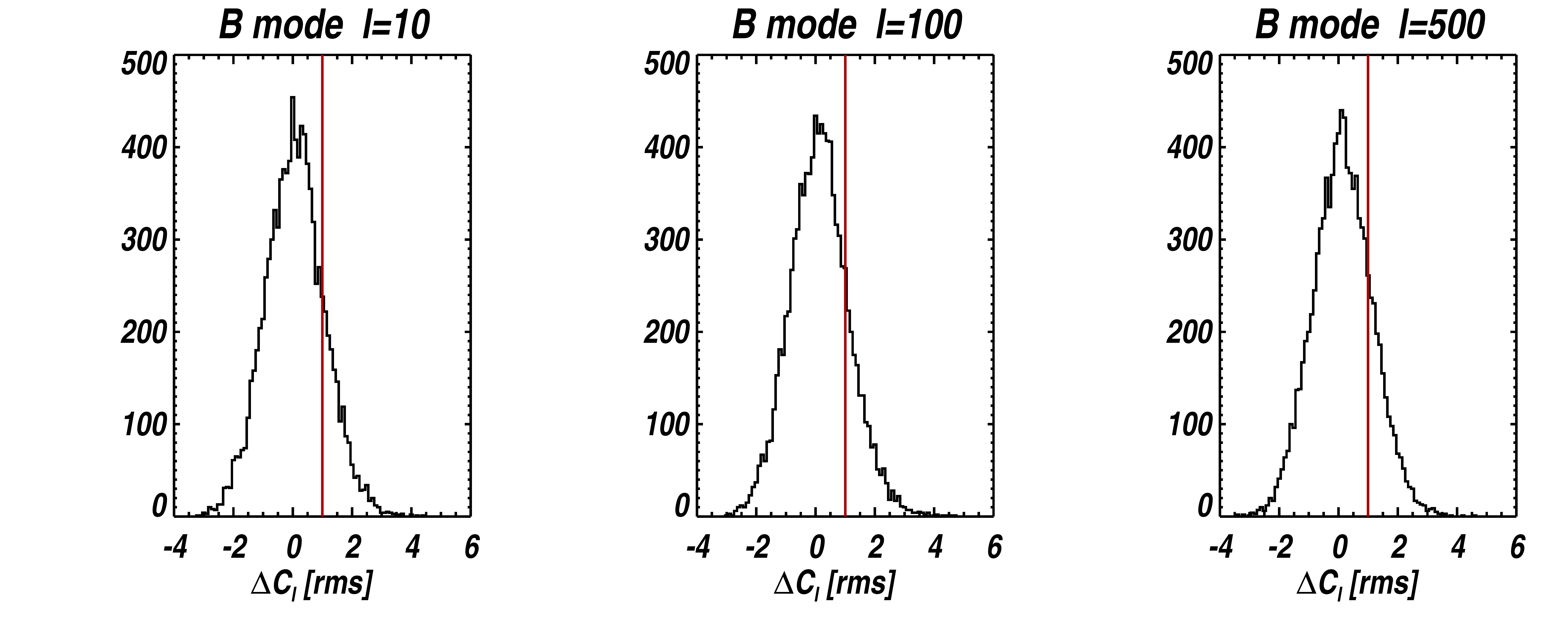}
	\caption{ \footnotesize Distribution of $\Delta C_\ell^{EE}$
          ({\it top}) and $\Delta C_\ell^{BB}$ ({\it bottom}) for
          $\sigma_\epsilon = 1\%$ polar efficiency errors for multipoles
          $\ell=10$, $\ell=100$, $\ell=500$, normalized to their rms
          (\textit{red line}).}
	\label{fig:distrib_efficiency}
	\end{center}
\end{figure}

\begin{figure}[!h]
	\begin{center}
	\includegraphics[width=9cm]{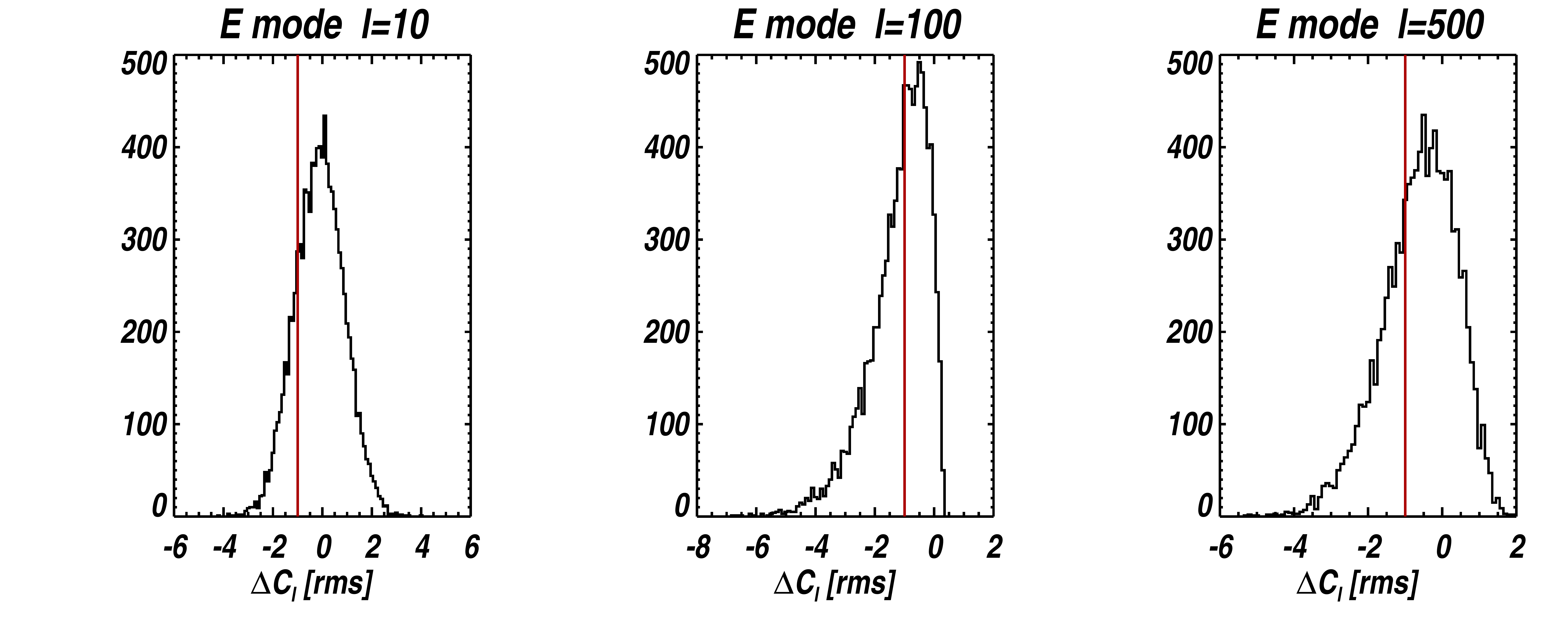}
	\includegraphics[width=9cm]{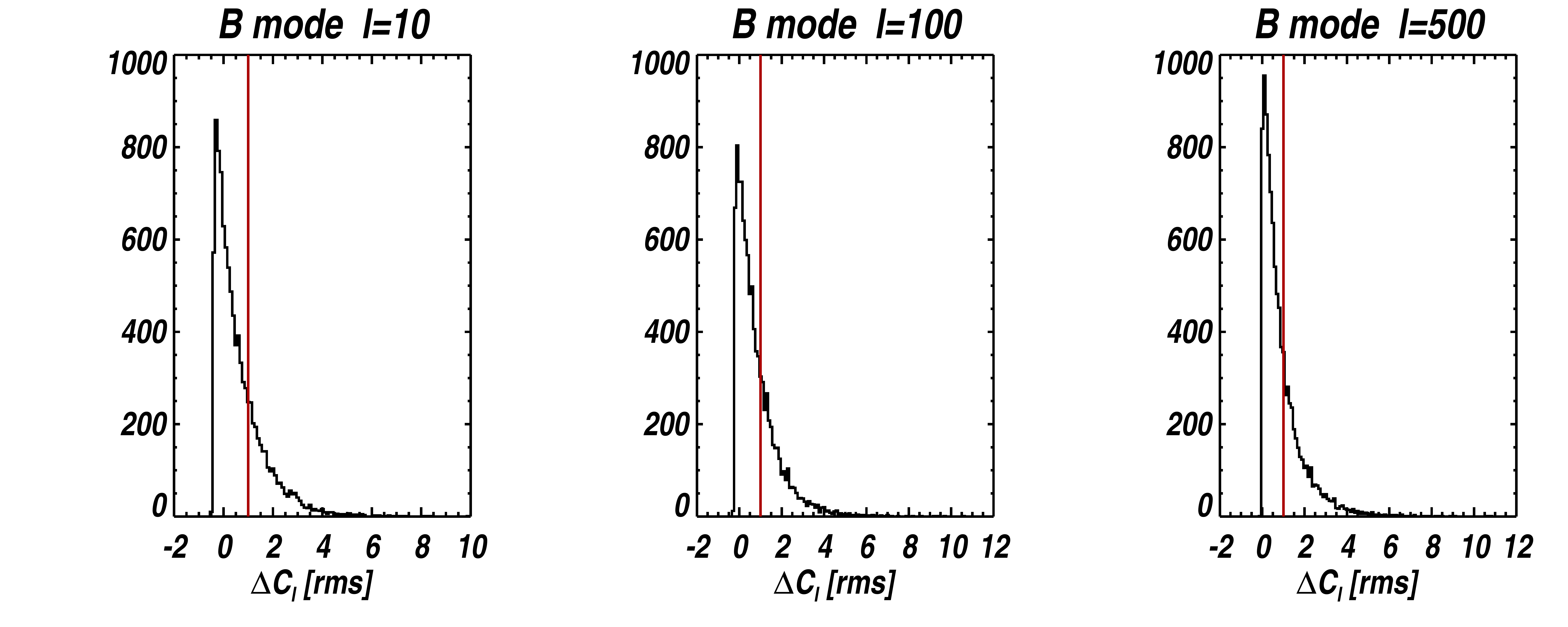}
	\caption{ \footnotesize Distribution of $\Delta C_\ell^{EE}$
          ({\it top}) and $\Delta C_\ell^{BB}$ ({\it bottom}) for
          $\sigma_\omega = 1^\circ$ orientation errors for multipoles
          $\ell=10$, $\ell=100$, $\ell=500$, normalized to their rms
          (\textit{red line}).}
	\label{fig:distrib_orientation}
	\end{center}
\end{figure}

The results show leakage coming from $TT$, $EE$ and $BB$
depending on the parameter considered. The gain uncertainty induces
leakage from intensity into polarization so $\Delta C_\ell$ show
leakage from $TT$ and $TE$ spectra (dominated by $TT$). For
polarization efficiency and orientation, $\Delta C_\ell$ is a
combination of $EE$ and $BB$ power spectra with relative weights that
depend on the distribution of uncertainties between the four bolometers
considered.  Due to second order terms, the distribution of
errors in the angular power spectra is highly non gaussian, as
shown in Figs.~\ref{fig:distrib_gain},~\ref{fig:distrib_efficiency}
and \ref{fig:distrib_orientation}.

\begin{figure}[hh!]
	\begin{center}
	\includegraphics[width=9cm]{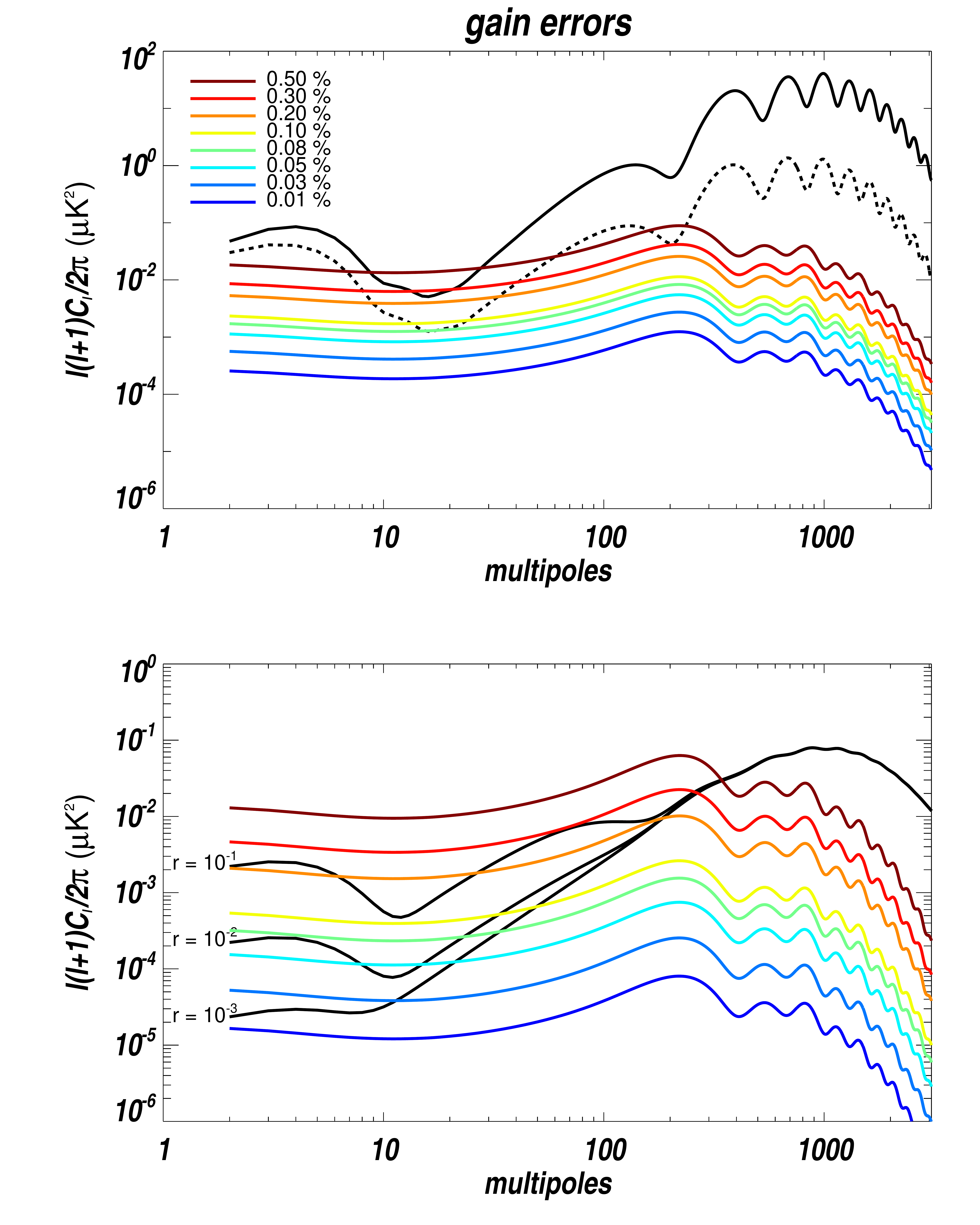}
	\caption{ \footnotesize $\Delta C_\ell$ in {\it rms} due to
          gain errors from 0.01\% to 1\% for $E$-mode ({\it top}) and
          $B$-mode ({\it bottom}) compared to initial spectrum ({\it
            solid black lines}). Cosmic variance for $E$-mode is
          plotted in {\it dashed black line}.}
	\label{fig:mc_gain}
	\end{center}
\end{figure}

\begin{figure}[hh!]
	\begin{center}
	\includegraphics[width=9cm]{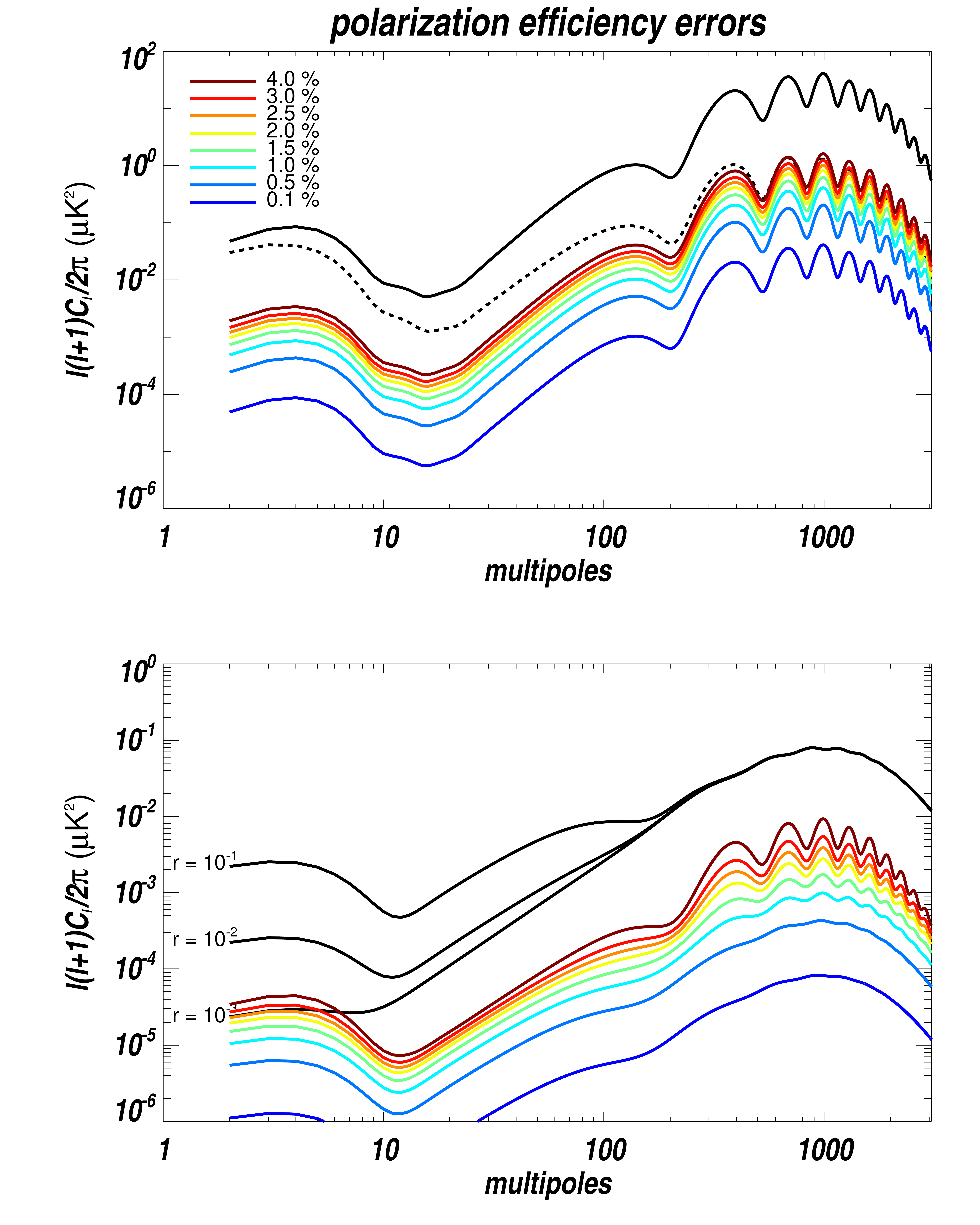}
	\caption{ \footnotesize $\Delta C_\ell$ in {\it rms} due to
          polarization efficiency errors from 0.1\% to 4\% for
          $E$-mode ({\it top}) and $B$-mode ({\it bottom}) compared to
          initial spectrum ({\it solid black lines}). Cosmic variance for
          $E$-mode is plotted in {\it dashed black line}.}
	\label{fig:mc_efficiency}
	\end{center}
\end{figure}

\begin{figure}[hh!]
	\begin{center}
	\includegraphics[width=9cm]{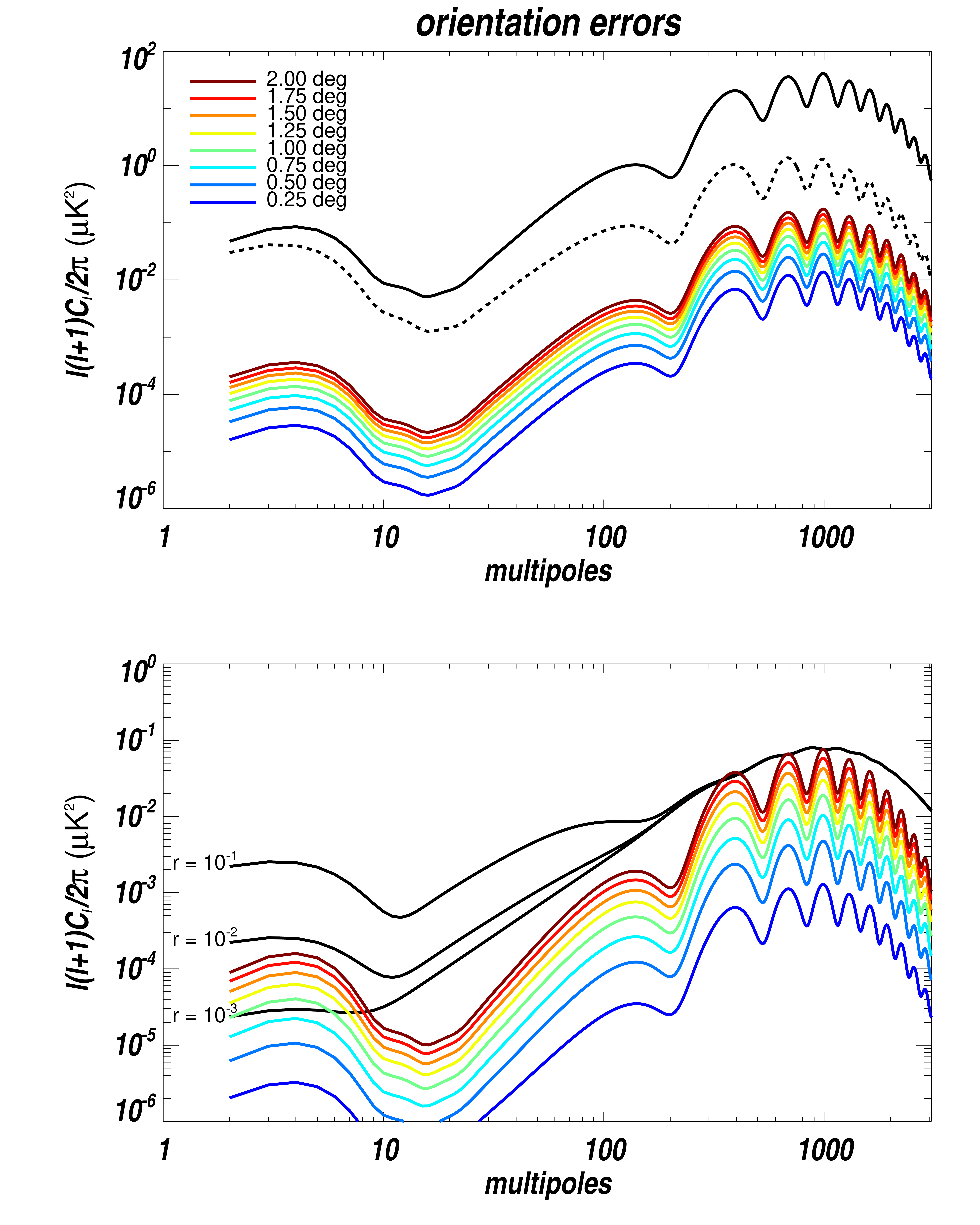}
	\caption{ \footnotesize $\Delta C_\ell$ in {\it rms} due to
          various orientation errors from 0.25 to 2 degrees for
          $E$-mode ({\it top}) and $B$-mode ({\it bottom}) compared to
          initial spectrum ({\it solid black lines}). Cosmic variance
          for $E$-mode is plotted in {\it dashed black line}.}
	\label{fig:mc_orientation}
	\end{center}
\end{figure}

We then compare the rms of those distributions for each multipole to
the cosmic variance of the $E$-mode and to an $r=0.05$ $B$-mode
spectrum with lensing (Figs.~\ref{fig:mc_gain},
\ref{fig:mc_efficiency} and \ref{fig:mc_orientation}).

Using these results, we can set the requirements for Planck-HFI on the
calibration of gain, polarization efficiency and orientation. More
precisely, we demand that the errors on the temperature and polarization
calibration parameters to be such that the induced leakage into the
$E$ power spectrum is lower than 10\% of the cosmic variance over the
multipole range $\ell = 2-1000$. This means that gains must be known to
$0.15\%$, polarization efficiencies to $0.5\%$ and detector
orientations to $1\deg$.

According to \cite{Efstathiou:2009kx}, an extended Planck mission
should be able to measure gravitational $B$-mode at a level of
$r=0.05$ and put an upper-limit of $r=0.03$ when considering
foreground residuals and noise levels. To achieve such a detection (or
upper-limit), the constraints on the calibration parameters must be
much tighter. For this goal, we set the leakage into $B$ power
spectrum to be 10\% of the $B$-mode model we want to target, for a
multipole range from $\ell=2$ to $100$. With such an hypothesis,
we find that the gain precision should be better than 0.05\% and the
orientations of the bolometers should be known to better than
0\pdeg75. The leakage due to polarization efficiency into B-mode is
very small (see bottom plot in Fig.~\ref{fig:mc_efficiency}), thus the
constraint on the polarization efficiency determination is not
relevant in that case (we found 10\%).

\section{Ground measurements}
\label{sec:ground_cal}

The Planck HFI polarization calibration on ground was divided into two
parts: polarization efficiencies were measured for each detector
separately, before focal plane assembly, at the University of Wales in
Cardiff in 2005, while orientations of the PSBs with respect to the
focal plane were measured during the overall calibration of the Planck
HFI in the Saturne cryostat at Orsay, France, in 2006.

\subsection{Polarization efficiency ground measurements}
\label{sec:xpolcal}

Detector-level polarization efficiency measurements were performed in
a 2-stage adiabatic demagnetization refrigerator (ADR) at a base
temperature of 200~mK.  The ADR was configured to take six detectors
per cooldown (in most cases all of the same optical band per
cooldown). Thermal blocking filters were used at the 4\,K, 77\,K and
300\,K stages of the testbed. The anti-reflective coating on the
cryostat window was matched to the optical band under test. The
window, of 125\,mm diameter, and all the thermal blockers were sized
such that they filled the beams. The polarization source was a
rotating polarizer grid positioned over an extended
temperature-controlled black body source of 75\,mm diameter running at
126$^\circ$C. The final source aperture was 70\,mm in diameter. The
mechanical structure of the source was fully clad with non-rotating
Eccosorb (type AN-72). The source was positioned approximately 690\,mm
from the cryostat window, tilted 4\pdeg8 off the optical axis, and
mechanically chopped at 6\,Hz. The experimental setup was fully
surrounded with Eccosorb (type AN-72) while the data were
recorded. Data were recorded in a step and sample fashion over five
full rotations of the polarizer grid with a 4\deg\ step size and a
4~second integration time.

Detailed results are given in the appendix in Tables~\ref{tab:cardpsb}
and~\ref{tab:cardswb} for PSBs and SWBs, respectively. The
polarization efficiency of the SWBs is low, as expected, and range between
1.6\% and 8.6\%. The statistical error is typically 0.5\%, and as much
as 1.8\% for one SWB. The polarization efficiency of the PSBs is typically
around 90\%, ranging from 84\% to 96\%, with errors below 0.3\%. 

\subsection{Orientation ground measurements}

\subsubsection{The calibration setup}

The orientation calibration was performed within a 1-meter diameter
cryostat cooled to 2\,K, to be close to flight conditions \citep[for a
  more detailed description of the calibration setup and photographs,
  see][]{Pajot:2010fr}. The detectors were cooled to their nominal
operating temperature, 100\,mK. For polarization measurements, the
source (Cold Source~2 or CS2) was a blackbody at 20\,K whose radiation
was diluted within a 50\,cm diameter sphere in order to illuminate,
after a reflection from mirror, the full focal plane at once. The
source was modulated by a diapason at a fixed frequency of 10\,Hz. The
radiation was linearly polarized by an aluminum grid deposited on a
138\,mm diameter mylar film. The aluminum strips of the polarizer were
5\microns\ wide, 5\microns\ thick and spaced 5\microns\ apart. The Mylar
film itself was 10\microns\ thick, with a transmission coefficient
greater than 0.9; the polarization efficiency of the polarizer was
measured to be better than 99.9\%, so it can be assumed equal to unity
at HFI frequencies. The polarizer could rotate freely around its
axis using a stepper motor. There are exactly 32,000 steps in one
rotation, so the precision in relative angle is better than
1\arcm.

\subsubsection{Reference for angle measurement}
\label{par:absangleprec}
The reference position was defined by a pin fixed to the polarizer,
which was detected by electric contact with a copper strip
with a precision of $\pm 5$ motor steps, {\it i.e.} $\pm$0\pdeg06. We
measured the angle of this reference position with respect to the
focal plane using the light of a laser diffracted by the strips of the
polarizer; the diffraction pattern is formed by points aligned
orthogonally to the strips ({\it i.e.}  parallel to the transmitted
polarization).

Two different methods were used to determine polarization angles with
respect to the focal plane. In the first method, we measured the
orientation with respect to the platform and used the mechanical position
of the instrument with respect to the platform to get the absolute
angle. In the second method, we measured the angle directly with
respect to the instrument. In both cases, we measured the same
angle and checked it was constant across the polarizer. Both methods
gave similar error estimates on the reference position angle, which
can safely be assumed to be lower than 0\pdeg3:
\begin{equation}
\left|\Delta\theta^{\mbox{\tiny absolute}}\right| < 0^\circ\!\!.3.
\end{equation}

\subsubsection{Data analysis}
\label{par:est_syst_err}

For this measurement, the polarizer was rotated by 5\deg\ steps and
signal was inegrated for 20\,s at each position. Eight full rotations
of the polarizer were performed.

At each polarizer position, the signal from the source is sinusoidal
with a frequency of 10\,Hz. It is demodulated fitting a sine curve over
a few periods, yielding around 60~independent measurements for each
stationary period of 20~seconds. The average and standard deviation of
these 60 measurements give the signal and its error for each 20~second
period, for a fixed position of the polarizer. The statistical error
was found to be typically below 1\% of the signal.

We then fit the signal as a function of the polarizer angle to
estimate the polarization efficiency and the orientation of the
detectors. However, despite the good quality of the polarizer, we
found cross-polarization leakage of around 30\%, much higher than that
found in Sect.~\ref{sec:xpolcal}, with the Cardiff measurements: it
was probably due to standing waves between the polarizer and the focal
plane and made the detector polarization efficiency unmeasurable with
this setup. The angle that maximizes the signal gives the orientation
of the polarizer; however, the PSB angle must be given in the horn
aperture plane, which is slightly out of parallel with the polarizer
plane. We have performed ray-tracing simulations to estimate and
correct for this geometrical effect. The corrections lie between
-0\pdeg5 to 0\pdeg5, and the precision (set by the precision on the
position of the polarizer) is better than 0\pdeg15.

\begin{figure}[tbp]
\begin{center}
	\includegraphics[width=9cm]{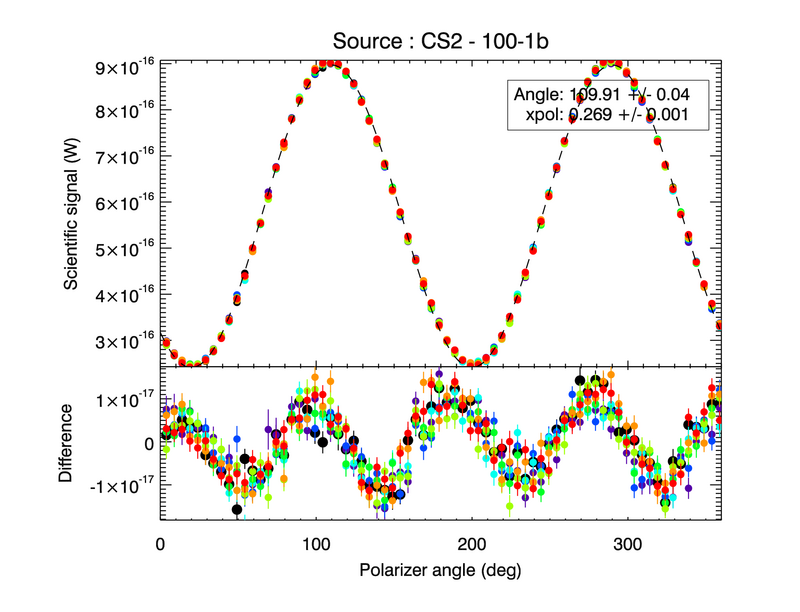}
\end{center}
\caption{ \footnotesize Signal of PSB 100-1b with respect to the
  angle in the horn aperture plane; each color represents one rotation
  of the polarizer (8 turns); the signal is fitted using a standard
  sine curve. The difference exhibits a systematic effect that can be
  explained by standing waves between the polarizer and the focal
  plane (see text).}
\label{fig:sig1001b}
\end{figure}

Figure~\ref{fig:sig1001b} shows the curve obtained for a PSB at
100\,GHz and the difference with the fitted model. The residuals show
a 90\deg-periodic sine curve, which is present in some detectors. Some
detectors also have glitches, reproduced at the same position at each
rotation of the polarizer. These glitches mostly affect the highest
two frequency channels (545 and 857\,GHz), i.e., only SWBs. As $\cos
2\theta$ and $\cos 4\theta$ are orthogonal functions over $2\pi$, the
fitted values for the angle and the polarization efficiency are
unchanged when adding such a term in the fitting model. However, we
cannot exclude that they may be contaminated by a systematic effect
like some other modes (mainly in mode $\cos 4\theta$). For example, if
the incoming radiation is the sum of two partially linearly polarized
radiations, one with orientation $\theta$ (rotating with the
polarizer) and one with fixed orientation $\theta_0$, the signal
measured by the detector reads:
\begin{eqnarray}
s(\theta) & \propto & 1 + \rho\cos 2(\theta-\theta_{\mbox{\tiny det}})
\nonumber \\ 
& & {}+ \rho^\prime\cos 2(\theta-\theta_0) \left[1 - \cos
  2(\theta-\theta_{\mbox{\tiny det}})\right]
\end{eqnarray}
where $\theta_{\mbox{\tiny det}}$ is the polarization orientation of
the detector. In this model, the angle measured through the phase of
the mode $\cos 2\theta$ will not be the detector polarization angle.

More generally, we can expand the signal as a Fourier series $s(\theta)
= \sum_{n=-N}^{+N}c_n \mbox{e}^{in\theta}$ and fit its coefficients
$c_n$ (which fulfill the condition $c_{-n}^\star = c_n$, as $s$ is a
real quantity). The coefficient $c_2$, giving the dependence in $\cos
2\theta$, contains the information on polarization efficiency and
angle through its modulus and argument, and is independent of the other
modes. To estimate the error on the polarization angle without relying
on a particular model, we assume that the mode $c_2$ is the sum of two
contributions, $c_2 = c_2^{\mbox{\tiny pol}} + c_2^{\mbox{\tiny
    syst}}$ (true polarization signal and induced systematic
effect). The maximum systematic error on the angle is then given by:
\begin{equation}
\max \left|\Delta\theta^{\mbox{\tiny syst}}\right| = \arctan
\left|\frac{c_2^{\mbox{\tiny syst}}}{c_2^{\mbox{\tiny pol}}}\right|.
\end{equation}
We draw an upper bound on the systematic error by assuming that
$|c_2^{\mbox{\tiny syst}}/c_2^{\mbox{\tiny pol}}| \lesssim \max_{n\neq
  0,2} |c_n/c_2|$. However, as the systematic error is due to complex
interference between the polarizer, the focal plane and the horns, we
chose a conservative limit by taking for all detectors the maximum of
this estimate among all PSBs. The statistical error on the
coefficients $c_n$ being negligible compared to the systematic error,
we finally find the following upper limit on the total error on the
relative angle of each polarization sensitive detectors:
\begin{equation}
\left|\Delta\theta^{\mbox{\tiny relative}}\right| < 0^\circ\!\!.9.
\end{equation}

As an independent check, we compared the relative angle between PSBs
within each horn (which is close but not exactly equal to 90\deg) with
the angles found using the setup described in
Sect.~\ref{sec:xpolcal}. We found an agreement within the systematic
error bars for all horns except one, which is, however, within the
statistical plus systematic error bar (the statistical error coming
from the Cardiff measurements).

The case of SWBs is treated separately, as the statistical error is
not negligible in this case (due to the low polarization
efficiencies). We performed a similar analysis, taking into account
the statistical error. The results are gathered in
table~\ref{tab:cardswb}. Note that the SWBs are not meant to be used
for polarization measurements.

\section{Discussion and conclusion}

This paper focuses on the impact of polarized calibration parameters
(gain, polarization efficiency and detector orientation) on power
spectra in the context of Planck-HFI. We have developed a
semi-analytical method that allows us to compute quickly and easily
the impact of uncertainties on gain, polarization efficiency and
orientation on the $E$ and $B$-mode power spectra, while exactly
accounting for the scanning strategy and the combination of different
detectors. We used this method in the particular case of Planck-HFI
and derived constraints on the gain, polarization efficiency and
detector orientation needed to achieve Planck-HFI's scientific goals.

Planck will use the orbital dipole to calibrate the total power for
each detector.  We find that the relative uncertainty on the gain must
be lower than 0.15\% to keep systematic error on $E$-mode power
spectrum below 10\% of the cosmic variance in the multipole range
$\ell = 2-1000$. Given the 0.2\% accuracy on relative gain obtained by
WMAP \citep{Hinshaw:2009os}, we expect that HFI can achieve the 0.15\%
requirements, thanks to the higher gain stability expected for HFI.

We show that the polarization efficiency uncertainty must be below
0.3\% in order to achieve the required sensitivity for the
$E$-mode. The error on the primordial $B$-mode power spectrum will be
kept below 10\% of the signal expected from a tensor-to-scalar ratio
$r=0.05$ in the multipole range $\ell=2-100$ if the polarization
efficiency is known to better than 10.3\%. In this paper, we have
presented the results of the ground measurements on HFI PSBs
polarization efficiency, which show an accuracy of 0.3\% that fulfills
the requirements for both $E$ and $B$-modes.

For the polarization orientation, we have distinguished a global
orientation error of the focal plane (which affects identically all
detectors) from a relative error (different for each detector). For
$E$-modes, we show that the requirement is 2\pdeg1 on the global
orientation knowledge and 1\deg\ on the relative orientation to keep
the error below 10\% of the cosmic variance in the range
$\ell=2-1000$. Both these requirements are already fulfilled by the
ground measurements, in which we found 0\pdeg3 and 0\pdeg9
respectively. In order to measure a $B$-mode signal with a systematic
error lower than 10\% for a tensor-to-scalar ratio $r=0.05$, the
global orientation must be known to better than 1\pdeg2 and the
relative orientation at better than 0\pdeg75. While the ground
measurements fulfill the requirement on global orientation, the
relative orientation knowledge will need to be improved in flight. For
Planck, we plan to use the Crab nebula as the primary polarization
calibrator \citep{Aumont:2009rc}, which will also allow the
  results presented in this paper to be cross-checked. The accuracy
of the ground measurements of polarization efficiencies and
orientations will allow the $E$-mode power spectrum to be measured,
with systematic errors lower than 10\% of the cosmic variance,
provided that the other sources of systematic effects are controlled.

\acknowledgement{Planck \emph{(http://www.esa.int/Planck)} is a
  project of the European Space Agency (ESA) with instruments provided
  by two scientific Consortia funded by ESA member states (in
  particular the lead countries: France and Italy) with contributions
  from NASA (USA), and telescope reflectors provided in a
  collaboration between ESA and a scientific Consortium led and funded
  by Denmark.}

\bibliographystyle{aa}


\onecolumn

\appendix
\section{Explicit forms of pointing related functions \label{se:appendix_deriv}}

We write the projection of the signal $\mathbf{m}$ into a sky map $\mathbf{s}$ as  
\begin{eqnarray}
	\hat{{\bf s}} 
	& = & 
	(\tilde{A}^T\tilde{A})^{-1}\tilde{A}^T{\bf m} \label{aeq:s_solve}
	\\
	& = & 
	\left[ \sum_d \tilde{A}_d^T\tilde{A}_d \right]^{-1} 
	\left[ \sum_d \tilde{A}_d^TA_d{\bf s} \right]
	\\
	& = & 
	\left[ \sum_d \tilde{A}_d^T\tilde{A}_d \right]^{-1} 
	\left[
		\sum_d \Lambda_d(\gamma_d, \epsilon_d, \omega_d){\bf s}
	\right].
\label{aeq:s}
\end{eqnarray}
Where $A$ is the pointing matrix. In this expression, $\Lambda_d(\gamma_d, \epsilon_d, \omega_d)$ is an explicit function of $\gamma_d$, $\epsilon_d$ and $\omega_d$. $\tilde g$, $\tilde{\rho}$ and $\tilde{\alpha}$ are only parameters.
If we note $t(d)$ the data samples of detector $d$, $\Lambda_d$ reads

\begin{equation}
	\Lambda_d = \sum_{t(d)} \left(
		\begin{array}{lll}
	 		(1+\gamma) & 
			 (\tilde\rho_d+\epsilon_\rho)\cos2(\tilde{\psi_d}(t)+\omega_d) & 
			 (\tilde\rho_d+\epsilon_\rho)\sin2(\tilde{\psi_d}(t)+\omega_d)
			 \\
			(1+\gamma) \tilde{\rho_d}\cos2\tilde{\psi_d}(t) &
			\tilde{\rho_d}(\tilde\rho_d+\epsilon_\rho)\cos2\tilde{\psi_d}(t)\cos2(\tilde{\psi_d}(t)+\omega_d) & 
			\tilde{\rho_d}(\tilde\rho_d+\epsilon_\rho)\cos2\tilde{\psi_d}(t)\sin2(\tilde{\psi_d}(t)+\omega_d) 
			\\
			(1+\gamma) \tilde{\rho_d}\sin2\tilde{\psi_d}(t) & 
			\tilde{\rho_d}(\tilde\rho_d+\epsilon_\rho)\sin2\tilde{\psi_d}(t)\cos2(\tilde{\psi_d}(t)+\omega_d) & 
			\tilde{\rho_d}(\tilde\rho_d+\epsilon_\rho)\sin2\tilde{\psi_d}(t)\sin2(\tilde{\psi_d}(t)+\omega_d)
		\end{array}
	\right) 
\end{equation}

Considering small variations around $\tilde{g}$, $\tilde{\rho}$ and $\tilde{\alpha}$, we can
write the perturbative expansion to first order for both $\gamma \ll 1$, $\epsilon \ll 1$
and $\omega \ll 1$ :
\begin{eqnarray}
	\Delta {\bf s} 
		& = & \hat{{\bf s}} - {\bf s} \\
		& = &
			\left[ \sum_d \tilde{A}_d^T \tilde{A}_d \right]^{-1}
			\left[ \sum_d \tilde{A}_d^T A_d - \tilde{A}_d^T \tilde{A}_d \right] {\bf s} \nonumber \\
		& \equiv & 
			\left[ \sum_d \tilde{A}_d^T \tilde{A}_d \right]^{-1}
			\left[ \sum_d \Lambda_d(\gamma_d, \epsilon_d, \omega_d) - \Lambda_d(0, 0, 0) \right] {\bf s} \nonumber \\
		& \simeq &
			\left[ \sum_d \tilde{A}_d^T \tilde{A}_d\right]^{-1}
			\sum_d \left[
				\frac{\partial \Lambda_d}{\partial\gamma_d} \gamma_d +
				\frac{\partial \Lambda_d}{\partial\epsilon_d} \epsilon_d +
				\frac{\partial \Lambda_d}{\partial\omega_d} \omega_d
			\right] {\mathbf s}
\label{aeq:delta_s}
\end{eqnarray}

Straightforward generalization to second order reads:

\begin{equation}
	\Delta {\bf s} = 
	\left[ \sum_d \tilde{A}_d^T \tilde{A}_d\right]^{-1}
  	\sum_d \left[
		\sum_{e\in\{\gamma,\epsilon,\omega\}} \frac{\partial \Lambda_d}{\partial e_d} e_d +
		\frac{1}{2}\sum_{(e,e')\in\{\gamma,\epsilon,\omega\}}\frac{\partial^2 \Lambda_d}{\partial e_d \partial e'_d} e_d e'_d
    \right] {\bf s}
\label{aeq:delta_s2}
\end{equation}

Derivatives of $\Lambda_d(\gamma_d, \epsilon_d, \omega_d)$ with respect to uncertainties of gain $\gamma$, polarization efficiency $\epsilon$ and detector orientation $\omega$ are given by
\begin{eqnarray}
	\left.\frac{\partial \Lambda_d}{\partial\gamma_d}\right|_{(0,0,0)} &= &
	\sum_{t(d)}
	\left(\begin{array}{lll}
		1 & 0 & 0\\
		\cos2\tilde{\psi}_d(t) & 0&0\\
		\sin2\tilde{\psi}_d(t) & 0 & 0\\
	\end{array}\right)
\label{aeq:dLdg} \\
	\left.\frac{\partial \Lambda_d}{\partial\epsilon_d}\right|_{(0,0,0)} &= &
	\sum_{t(d)}
	\left(\begin{array}{lll}
		0 & \cos2\tilde{\psi}_d(t) & \sin2\tilde{\psi}_d(t)\\
		0 & \tilde{\rho_d}\cos^22\tilde{\psi}_d(t) & \tilde{\rho_d}\cos2\tilde{\psi}_d(t)\sin2\tilde{\psi}_d(t) \\
		0 & \tilde{\rho_d}\cos2\tilde{\psi}_d(t)\sin2\tilde{\psi}_d(t) & \tilde{\rho_d}\sin^22\tilde{\psi}_d(t)
	\end{array}\right)
\label{aeq:dLdr} \\
	\left.\frac{\partial \Lambda_d}{\partial \omega_d}\right|_{(0,0,0)} & = & 
	\sum_{t(d)}
	\left(\begin{array}{lll}
		  0 & -2\tilde{\rho_d}\sin2\tilde{\psi}_d(t) & 2\tilde{\rho_d}\cos2\tilde{\psi}_d(t)\\
		  0 & -2\tilde{\rho_d}^2\cos2\tilde{\psi}_d(t)\sin2\tilde{\psi}_d(t) & 2\tilde{\rho_d}^2\cos^22\tilde{\psi}_d(t)\\
		  0 & -2\tilde{\rho_d}^2\sin^22\tilde{\psi}_d(t) & 2\tilde{\rho_d}^2\cos2\tilde{\psi}_d(t)\sin2\tilde{\psi}_d(t)
	\end{array}\right)
\label{aeq:dLde}
\end{eqnarray}
And second order derivatives reads
\begin{eqnarray}
	\left.\frac{\partial^2 \Lambda_d}{\partial \gamma_d^2}\right|_{(0,0,0)} & = & 0
\label{aeq:d2Ldg2}  \\
	\left.\frac{\partial^2 \Lambda_d}{\partial \epsilon_d^2}\right|_{(0,0,0)} & = & 0
\label{aeq:d2Ldr2}  \\
	\left.\frac{\partial^2 \Lambda_d}{\partial \omega_d^2}\right|_{(0,0,0)} & = & 
	\sum_{t(d)}
	\left(\begin{array}{lll}
		  0 & -4\tilde{\rho_d}\cos2\tilde{\psi}_d(t) & -4\tilde{\rho_d}\sin2\tilde{\psi}_d(t)\\
		  0 & -4\tilde{\rho_d}^2\cos^22\tilde{\psi}_d(t) & -4\tilde{\rho_d}^2\cos2\tilde{\psi}_d(t)\sin2\tilde{\psi}_d(t)\\
		  0 & -4\tilde{\rho_d}^2\cos2\tilde{\psi}_d(t)\sin2\tilde{\psi}_d(t) & -4\tilde{\rho_d}^2\sin^22\tilde{\psi}_d(t)
	\end{array}\right)
\label{aeq:d2Lde2}  \\
	\left.\frac{\partial^2 \Lambda_d}{\partial \epsilon_d\partial \omega_d}\right|_{(0,0,0)} & = & 
	\sum_{t(d)}
	\left(\begin{array}{lll}
		  0 & -2\sin2\tilde{\psi}_d(t) & 2\cos2\tilde{\psi}_d(t)\\
		  0 & -4\tilde{\rho_d}\cos2\tilde{\psi}_d(t)\sin2\tilde{\psi}_d(t) & 4\tilde{\rho_d}\cos^22\tilde{\psi}_d(t)\\
		  0 & -4\tilde{\rho_d}\sin^22\tilde{\psi}_d(t) & 4\tilde{\rho_d}\cos2\tilde{\psi}_d(t)\sin2\tilde{\psi}_d(t)
	\end{array}\right)
\label{aeq:d2Ldrde}  \\
	\left.\frac{\partial^2 \Lambda_d}{\partial \gamma_d\partial \epsilon_d}\right|_{(0,0,0)} & = & \left.\frac{\partial^2 \Lambda_d}{\partial \gamma_d\partial \omega_d}\right|_{(0,0,0)} = 0
\label{aeq:d2Ldgde}  
\end{eqnarray}

\newpage

\section{Polarization efficiencies and angles}

\begin{table}[!hh]
\tiny
 \newdimen\digitwidth 
  \setbox0=\hbox{\rm 0} 
  \digitwidth=\wd0 
  \catcode`*=\active 
  \def*{\kern\digitwidth} 
  \caption{\label{tab:cardpsb}Polarization efficiencies and orientations for Planck-HFI PSBs}
  \begin{center}
    \begin{tabular}{ccc}
      \hline\hline
      Bolometer (PSB) \T\B & Polarization efficiency [\%] &
    Polarization angle \\
      \hline
\T  100-1a &  94.7 $\pm$   0.2 & *21\pdeg1 $\pm$ 0\pdeg9 [rel] $\pm$ 0\pdeg3 [abs]\\
    100-1b &  94.3 $\pm$   0.3 & 109\pdeg9 $\pm$ 0\pdeg9 [rel] $\pm$ 0\pdeg3 [abs]\\
    100-2a &  96.2 $\pm$   0.2 & *44\pdeg3 $\pm$ 0\pdeg9 [rel] $\pm$ 0\pdeg3 [abs]\\
    100-2b &  90.2 $\pm$   0.2 & 133\pdeg5 $\pm$ 0\pdeg9 [rel] $\pm$ 0\pdeg3 [abs]\\
    100-3a &  90.1 $\pm$   0.3 & **0\pdeg7 $\pm$ 0\pdeg9 [rel] $\pm$ 0\pdeg3 [abs]\\
    100-3b &  93.4 $\pm$   0.2 & *90\pdeg6 $\pm$ 0\pdeg9 [rel] $\pm$ 0\pdeg3 [abs]\\
    100-4a &  95.7 $\pm$   0.3 & 158\pdeg5 $\pm$ 0\pdeg9 [rel] $\pm$ 0\pdeg3 [abs]\\
    100-4b &  92.3 $\pm$   0.2 & *70\pdeg0 $\pm$ 0\pdeg9 [rel] $\pm$ 0\pdeg3 [abs]\\
    143-1a &  83.3 $\pm$   0.2 & *42\pdeg9 $\pm$ 0\pdeg9 [rel] $\pm$ 0\pdeg3 [abs]\\
    143-1b &  84.6 $\pm$   0.2 & 135\pdeg2 $\pm$ 0\pdeg9 [rel] $\pm$ 0\pdeg3 [abs]\\
    143-2a &  87.5 $\pm$   0.3 & *44\pdeg2 $\pm$ 0\pdeg9 [rel] $\pm$ 0\pdeg3 [abs]\\
    143-2b &  89.3 $\pm$   0.3 & 134\pdeg0 $\pm$ 0\pdeg9 [rel] $\pm$ 0\pdeg3 [abs]\\
    143-3a &  83.9 $\pm$   0.2 & **0\pdeg4 $\pm$ 0\pdeg9 [rel] $\pm$ 0\pdeg3 [abs]\\
    143-3b &  89.9 $\pm$   0.2 & *93\pdeg7 $\pm$ 0\pdeg9 [rel] $\pm$ 0\pdeg3 [abs]\\
    143-4a &  93.1 $\pm$   0.2 & **3\pdeg1 $\pm$ 0\pdeg9 [rel] $\pm$ 0\pdeg3 [abs]\\
    143-4b &  92.8 $\pm$   0.2 & *91\pdeg5 $\pm$ 0\pdeg9 [rel] $\pm$ 0\pdeg3 [abs]\\
    217-5a &  95.0 $\pm$   0.1 & *44\pdeg7 $\pm$ 0\pdeg9 [rel] $\pm$ 0\pdeg3 [abs]\\
    217-5b &  95.2 $\pm$   0.2 & 133\pdeg9 $\pm$ 0\pdeg9 [rel] $\pm$ 0\pdeg3 [abs]\\
    217-6a &  94.9 $\pm$   0.2 & *45\pdeg0 $\pm$ 0\pdeg9 [rel] $\pm$ 0\pdeg3 [abs]\\
    217-6b &  95.4 $\pm$   0.2 & 134\pdeg8 $\pm$ 0\pdeg9 [rel] $\pm$ 0\pdeg3 [abs]\\
    217-7a &  94.0 $\pm$   0.2 & **0\pdeg3 $\pm$ 0\pdeg9 [rel] $\pm$ 0\pdeg3 [abs]\\
    217-7b &  93.7 $\pm$   0.1 & *91\pdeg2 $\pm$ 0\pdeg9 [rel] $\pm$ 0\pdeg3 [abs]\\
    217-8a &  94.2 $\pm$   0.1 & **2\pdeg2 $\pm$ 0\pdeg9 [rel] $\pm$ 0\pdeg3 [abs]\\
    217-8b &  94.1 $\pm$   0.1 & *92\pdeg5 $\pm$ 0\pdeg9 [rel] $\pm$ 0\pdeg3 [abs]\\
    353-3a &  88.7 $\pm$   0.1 & *44\pdeg1 $\pm$ 0\pdeg9 [rel] $\pm$ 0\pdeg3 [abs]\\
    353-3b &  92.0 $\pm$   0.1 & 132\pdeg4 $\pm$ 0\pdeg9 [rel] $\pm$ 0\pdeg3 [abs]\\
    353-4a &  87.0 $\pm$   0.1 & *45\pdeg3 $\pm$ 0\pdeg9 [rel] $\pm$ 0\pdeg3 [abs]\\
    353-4b &  91.4 $\pm$   0.1 & 135\pdeg2 $\pm$ 0\pdeg9 [rel] $\pm$ 0\pdeg3 [abs]\\
    353-5a &  84.4 $\pm$   0.1 & 178\pdeg4 $\pm$ 0\pdeg9 [rel] $\pm$ 0\pdeg3 [abs]\\
    353-5b &  87.4 $\pm$   0.1 & *90\pdeg3 $\pm$ 0\pdeg9 [rel] $\pm$ 0\pdeg3 [abs]\\
    353-6a &  87.3 $\pm$   0.1 & **1\pdeg3 $\pm$ 0\pdeg9 [rel] $\pm$ 0\pdeg3 [abs]\\
    353-6b &  88.5 $\pm$   0.1 & *91\pdeg2 $\pm$ 0\pdeg9 [rel] $\pm$ 0\pdeg3 [abs]\\
      \hline
    \end{tabular}
  \end{center}
\tablefoot{Ideal PSBs should have a
  100\% polarization efficiency. The error on polarization efficiency
  is only statistical. Error on polarization orientation is due to
  systematics: the absolute error is due to the error on the
  measurement of the reference position; the relative error is due to
  an optical systematic effect in the Saturne cryostat. The
  statistical errors are negligible and therefore not shown in this table.}
\end{table}

\begin{table}[!hh]
\tiny
 \newdimen\digitwidth 
  \setbox0=\hbox{\rm 0} 
  \digitwidth=\wd0 
  \catcode`*=\active 
  \def*{\kern\digitwidth} 
  \caption{\label{tab:cardswb}Polarization efficiencies and
  orientations for Planck-HFI SWBs}
  \begin{center}
    \begin{tabular}{ccc}
      \hline\hline
      Bolometer (SWB) \T\B &  Polarization efficiency [\%]  &
    Polarization angle \\
      \hline
\T 143-5 &  6.6  $\pm$ 0.3 & *65\pdeg7 $\pm$  0\pdeg1 [stat] $\pm$ *0\pdeg6 [syst]   \\
   143-6 &  4.4  $\pm$ 0.3 & *70\pdeg6 $\pm$  0\pdeg2 [stat] $\pm$ *4\pdeg7 [syst]   \\
   143-7 &  1.7  $\pm$ 0.4 & 102\pdeg8 $\pm$  0\pdeg2 [stat] $\pm$ *1\pdeg7 [syst]  \\
   143-8 &  1.6  $\pm$ 0.5 & *75\pdeg7 $\pm$  0\pdeg3 [stat] $\pm$ *4\pdeg4 [syst]   \\
   217-1 &  4.0  $\pm$ 0.2 & *98\pdeg4 $\pm$  2\pdeg3 [stat] $\pm$ *5\pdeg5 [syst]   \\
   217-2 &  2.1  $\pm$ 0.1 & *82\pdeg5 $\pm$  1\pdeg5 [stat] $\pm$ *4\pdeg9 [syst]   \\
   217-3 &  4.1  $\pm$ 0.2 & 170\pdeg9 $\pm$  0\pdeg9 [stat] $\pm$ *2\pdeg1 [syst]  \\
   217-4 &  3.8  $\pm$ 0.6 & 120\pdeg0 $\pm$  1\pdeg2 [stat] $\pm$ *2\pdeg7 [syst]  \\
   353-1 &  3.4  $\pm$ 0.2 & 103\pdeg1 $\pm$  1\pdeg2 [stat] $\pm$ *3\pdeg6 [syst]  \\
   353-2 &  4.8  $\pm$ 0.1 & 114\pdeg6 $\pm$  0\pdeg5 [stat] $\pm$ *2\pdeg7 [syst]  \\
   353-7 &  8.1  $\pm$ 0.1 & 121\pdeg5 $\pm$  0\pdeg8 [stat] $\pm$ *4\pdeg2 [syst]  \\
   353-8 &  7.9  $\pm$ 0.1 & 133\pdeg0 $\pm$  0\pdeg3 [stat] $\pm$ *1\pdeg9 [syst]  \\
   545-1 &  4.7  $\pm$ 0.1 & 129\pdeg1 $\pm$  1\pdeg0 [stat] $\pm$ *2\pdeg4 [syst]  \\
   545-2 &  5.7  $\pm$ 0.1 & 139\pdeg1 $\pm$  0\pdeg7 [stat] $\pm$ *1\pdeg3 [syst]  \\
   545-3 &  5.3  $\pm$ 0.1 & 150\pdeg3 $\pm$  0\pdeg8 [stat] $\pm$ *2\pdeg4 [syst]  \\
   545-4 &  5.9  $\pm$ 0.1 & 145\pdeg6 $\pm$  0\pdeg8 [stat] $\pm$ *1\pdeg7 [syst]  \\
   857-1 &  7.8  $\pm$ 1.8 & 157\pdeg3 $\pm$  2\pdeg1 [stat] $\pm$ *5\pdeg1 [syst]  \\
   857-2 &  6.3  $\pm$ 0.1 & 108\pdeg4 $\pm$  4\pdeg0 [stat] $\pm$ 16\pdeg5 [syst]  \\
   857-3 &  8.6  $\pm$ 0.8 & 176\pdeg8 $\pm$  1\pdeg4 [stat] $\pm$ *2\pdeg6 [syst]  \\
   857-4 &  6.3  $\pm$ 0.8 & 161\pdeg9 $\pm$  2\pdeg3 [stat] $\pm$ *6\pdeg2 [syst]  \\
      \hline
    \end{tabular}
  \end{center}
\tablefoot{Ideal SWBs should have a null polarization efficiency. Global uncertainty (0\pdeg3) 
   is common for all detector and not added.}
\end{table}

\end{document}